\documentclass[useAMS,usegraphicx]{mn2e}
\usepackage{amsmath}

\newcommand{\AAA}[3]    {\mbox{#3, A\&A,~#1,~#2}}

\newcommand{\ApJ}[3]    {\mbox{#3, ApJ,~#1,~#2}}

\newcommand{\ARAA}[3]   {\mbox{#3, Ann.~Rev.~A~\&~A,~#1,~#2}}

\newcommand{\MNRAS}[3]  {\mbox{#3, MNRAS,~#1,~#2}}
\newcommand{\Nature}[3] {\mbox{#3, Nature,~#1,~#2}}

\def\ap3m{AP$^3$M}

\def\h0{$h_0$}
\def\H0{$H_0$}

\def\Msun{${\rm M}_{\odot}~$}

\def\d{{\rm ~d}}

\def\ln{{~\rm ln}}

\def\ea{et~al.~} 

\def\lesssim{\mathrel{\hbox{\rlap{\hbox{\lower4pt\hbox{$\sim$}}}\hbox{$<$}}}}
\def\grtsim{\mathrel{\hbox{\rlap{\hbox{\lower4pt\hbox{$\sim$}}}\hbox{$>$}}}}

\newcommand{\bh}    {\bullet}

\renewcommand{\r}       {\tilde{r}}

\begin{document}
\title[MBH remnants of the first stars II: optical and X-ray signatures
in present-day galactic haloes]{Massive black hole remnants of the
  first stars II: optical and X-ray signatures
in present-day galactic haloes}

\author[Islam R.R., Taylor J.E. \& Silk J.]
       {Ranty R. Islam\thanks{Email: rri@astro.ox.ac.uk}, James E. Taylor\thanks{PPARC fellow} and Joseph Silk\\
       {Astrophysics, Denys Wilkinson Building, Keble Road, Oxford, OX1 3RH, UK}}

\date{Accepted 2004 July 7. Received 2004 July 7; in original form 2003 October 24}

\maketitle

\begin{abstract}
The first stars forming in minihaloes at redshifts greater than 20 may
have been very massive, and could have left behind massive black hole (MBH)
remnants. In a previous paper we investigated the hierarchical merging
of these MBHs and their associated haloes, using a semi-analytical
approach consisting of a hierarchical merger tree algorithm and
explicit prescriptions for the dynamics of merged substructure inside a
larger host halo following a merger.
One of the results was the prediction of a number of MBHs orbiting
throughout present-day galactic haloes. In addition, we estimated the
mass-accretion rate of these MBHs, assuming that they retained
around them a core of material from the original haloes in which they formed.
On the basis of these estimates, in this paper we determine
the bolometric, optical and X-ray luminosity functions for
accreting MBHs, using thin disk and advection dominated accretion flow models.
Our predicted MBH X-ray fluxes are then compared with observations of
ultra-luminous X-ray sources in galaxies. We find that the slope
and normalisation of the predicted X-ray luminosity functions are
similar to those observed, suggesting that MBHs could account for
some fraction of these sources.
\end{abstract}

\section{Introduction}
There is increasing speculation that the first generation of stars
in the universe may have been extremely massive, and that some of
these objects could have collapsed directly to massive black holes (MBHs)
at the end of a brief stellar lifetime. If this is indeed the case,
than it has interesting implications for the formation of
the super-massive black holes (SMBHs) and unusual X-ray sources
observed in the local universe.

Recent semi-analytic \cite{hutchings02,fuller00,tegmark97}
and numerical studies \cite{bromm02,abel00} suggest that the
first stars in the universe formed inside dense baryonic cores,
as they cooled and collapsed within dark matter haloes at very high redshift.
For a standard $\Lambda$CDM cosmology, these {\em minihaloes} are
estimated to have had masses in the range
$M_{min} \sim 10^5$--$10^6 ~h^{-1}$ \Msun,
and to have collapsed at redshifts $z_{\rm collapse} \sim 20$--$30$
or higher.
Since these first star-forming clouds contained essentially no metals,
gas cooling would proceed much more slowly in these systems
than in present-day molecular clouds. As a result, they may have
collapsed smoothly and without fragmentation, producing dense cores much
more massive than the proto-stellar cores observed in star-formation
regions today. Assuming nuclei within these cores accreted the surrounding
material efficiently, the result would be a first generation of protostars
with masses as great as $10^3$ \Msun \cite{bromm02,omukai01}.

As yet, nothing definite is known about the subsequent evolution
of such objects. However, their large masses would probably result
in many of them ending up as MBHs with little intervening
mass loss -- for systems in this mass range, gravity is so strong
that there is no ejection of material from the system in a final
supernova bounce \cite{heger02}.
This high-redshift population of MBHs is particularly interesting,
since it might seed the formation of the SMBHs seen at the centres
of galaxies at the present-day.

In a previous paper \cite{islam_I} we used a semi-analytical model
of galaxy formation to follow the evolution of MBHs as they
merge together hierarchically, along with their associated haloes.
The code we used combines a Monte-Carlo algorithm to generate halo
merger trees with analytical descriptions for the main dynamical processes --
dynamical friction, tidal stripping, and tidal heating --
that determine the evolution of merged remnants within a galaxy halo.
We introduce seeds into the code by assuming that in each minihalo
forming before a redshift $z_{\rm collapse}$, a single  MBH forms as the
end result of primordial star formation.

For our computations, we considered four different sets of values for
the parameters $\nu_{pk}$ and $M_{\bh,seed}$, which fix the abundance
and the mass of the seeds respectively. These values are summarised in
table \ref{tab:inits}.
Our choice of a seed MBH mass of 260 \Msun is motivated by the result
of Heger (2002) \nocite{heger02} that massive stars above this mass
will not experience a supernova at the end of their lives, but instead
will collapse directly to a MBH of essentially the same mass.
\begin{table}
  \begin{center}
    \caption{The mass of seed MBHs, and the height of the peaks in the initial
    density field in which they formed, for the four models considered.
    The collapse redshift $z_{\rm collapse}$ is
    the epoch when peaks of height $\nu_{pk}$ first
    cross the cooling threshold (see paper I).}
    \begin{tabular}{cccc} \\    \hline \hline \label{tab:inits}
      Model & $M_{\bh,seed}$ & peak height $\nu_{pk}$ & $z_{\rm collapse}$\\ \hline
      A & 260 \Msun & 3.0 & 24.6\\
      B & 1300 \Msun & 3.0&  24.6\\
      C & 260 \Msun & 2.5 & 19.8\\
      D & 260 \Msun & 3.5 & 29.4\\ \hline \hline
    \end{tabular}
  \end{center}
\end{table}

\begin{figure}
  \centering
  \includegraphics[width=84mm]{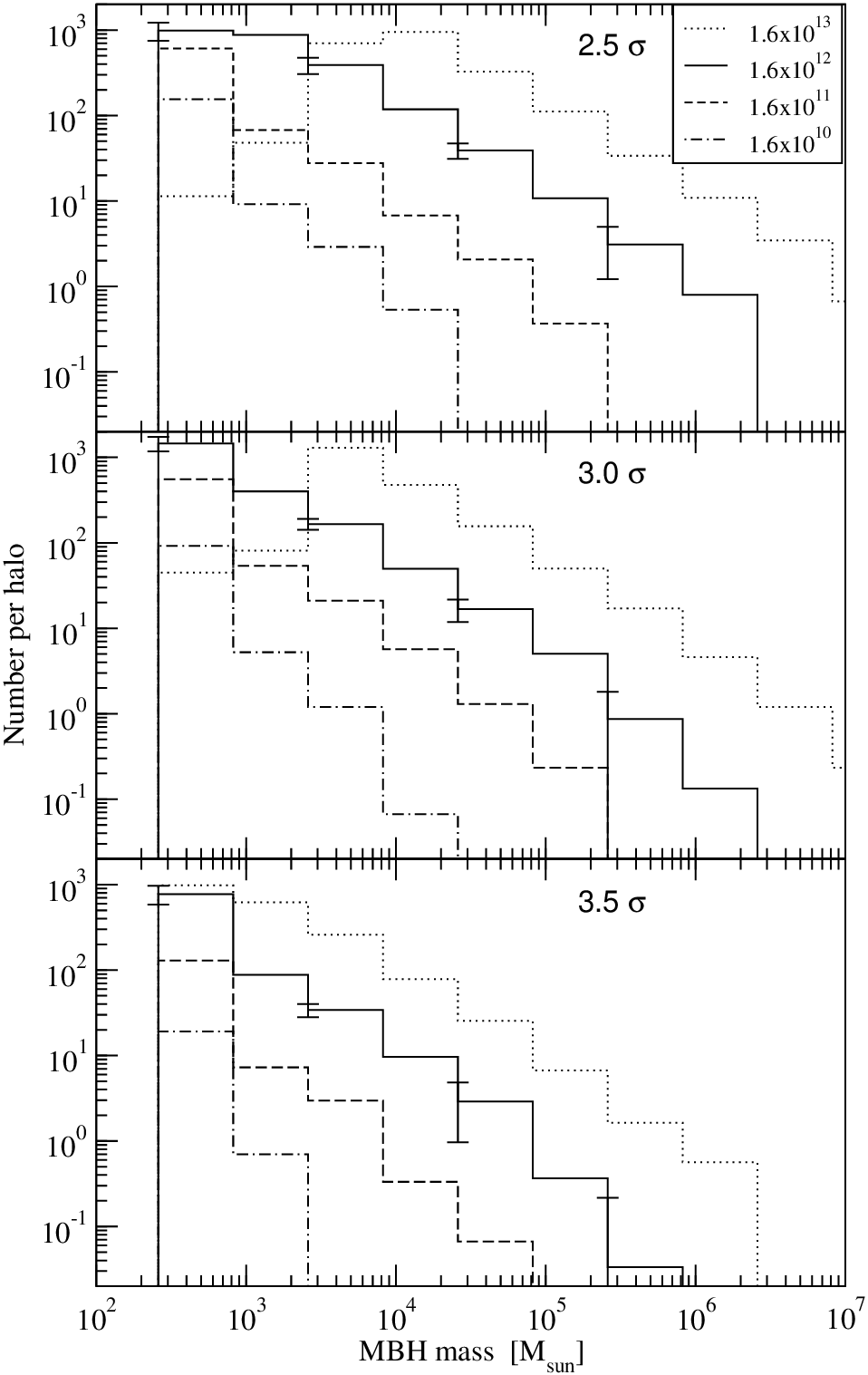}
  \caption{Abundance of all MBHs in the halo for models C, A and D (top to
    bottom panels) for different final halo masses averaged over 30 trees with error bars corresponding
    to the standard deviation.}
  \label{fig:MBHmassfunc}
\end{figure}

In paper I we investigated the abundance of MBHs in present-day galaxies
as a result of this process of hierarchical merging and dynamical
evolution. This is shown in figure \ref{fig:MBHmassfunc}.
We now look at how this model could be tested observationally in nearby
galaxies.
As the  MBHs orbit they will emit in the optical, UV and X-ray due to
accretion of matter in their direct
vicinity. This matter can be either the interstellar medium (ISM) of the
host, or remnant material of the original satellite that remains bound
to the MBH.
While optical and UV emissions could in principle arise from a variety
of sources and be subject to a range of absorption and distortion
mechanisms, X-ray emissions are more uniquely associated with
accreting black holes. In addition, hard X-rays are much less subject
to absorption. Observations of luminous and ultra-luminous X-ray point
sources associated with galactic haloes in the local universe might
therefore provide constraints on the existence and abundance of a
population of MBHs. Optical and UV observations could be used as
a follow up, to confirm the case for detection of a MBH.

In section \ref{sec:accretionprocesses} we outline the accretion
scenarios and mechanisms we considered. These include
thin disk and advection dominated accretion flows in particular. In section
\ref{sec:lum_functions} we present the resulting optical and X-ray
luminosity functions for MBHs orbiting in present-day haloes.
This is then compared with observations of ultraluminous X-ray sources (ULXs)
in section \ref{sec:observations}. We sum up our results and conclude
in section \ref{sec:summary}.

\section{Accretion Processes}
\label{sec:accretionprocesses}
\subsection{Accretion rates}
We assume that accretion onto MBHs orbiting in present-day haloes is
given by the Bondi-Hoyle mass-accretion rate \cite{bondi44,bondi52}
\begin{equation}
    \frac{\d M}{\d t} = \pi r_{acc}^2 \sqrt{v_{\bullet}^2 + c_s^2} \rho_g
\end{equation}
Here $c_s$ is the sound speed in the gas and $\rho_g$ its density -- both far from the MBH. $v_{\bullet}$ is the
velocity of the MBH and $r_{acc}$ the accretion radius
\begin{equation}
    r_{acc} = \frac{2 G M_{\bullet}}{v_{\bullet}^2 + c_s^2}
\end{equation}
giving an accretion rate
\begin{equation}
    \frac{\d M}{\d t} = \frac{4 \pi G^2 M_{\bullet}^2 \rho_g}{c_s^3} ( 1 + \beta_s^2)^{-3/2}
\end{equation}
where we have used $\beta_s \equiv v_{\bullet}/c_s$ \cite{chisholm02}. If the MBH accretes adiabatically from a
gas of pure hydrogen this is
\begin{eqnarray}
    \dot{M} &=& 8.77 \times 10^{-12} \left(\frac{M_{\bullet}}{100
M_{\odot}}\right)^2 \nonumber \left(\frac{\rho_g}{10^{-24} {\rm g
~cm}^{-3}}\right) \\
& & \times \left(\frac{c_s}{10 {\rm ~km ~s}^{-1}}\right)^{-3} (1 + \beta_s^2)^{-3/2} ~M_{\odot} {\rm yr}^{-1}
\end{eqnarray}
If a constant fraction $\eta$ of the accreted mass is radiated away,
the bolometric accretion luminosity of the MBH is then
\begin{eqnarray} \label{eq:Lbol}
    L_{bol} &=& \eta \frac{dM}{dt} c^2 \nonumber \\
&=& \eta ~5\times 10^{35} \left(\frac{M_{\bh}}{100
M_{\odot}}\right)^2 \left(\frac{\rho_g}{10^{-24} {\rm ~g
~cm}^{-3}}\right) \nonumber \\
& & \times \left(\frac{c_s}{10 {\rm ~km ~s}^{-1}}\right)^{-3} (1 +
\beta_s^2)^{-3/2} {\rm ~erg ~s}^{-1}
\end{eqnarray}
where $c$ is the speed of light.

In what follows we assume that while the nature of the accretion process
is somewhat uncertain, the overall mass-accretion rate is essentially
determined by the Bondi-Hoyle formula. We neglect the possibility that
the mass-accretion rate is modified e.g. by a non-negligible mass of
an accretion disk that may form around the accreting MBHs.

As the MBHs orbit through the host halo they accrete matter from the
host ISM. In paper I we have shown that ISM accretion rates are too
small to lead to significant accretion signatures even in regions
with relatively large amounts of gas, such as the disk or the galactic
centre. Much higher rates are possible if MBHs accrete from a residual
core of material from the satellite they were originally associated with.
Such a core, consisting of gas and possibly stars, could remain bound even
if the original satellite has lost all but a few percent of its initial mass.
While insignificant for the overall dynamics of the satellite, the
amount of baryonic matter contained in this core could still contribute
significantly to the accretion onto a MBH that is present at its centre and so
boost its accretion luminosity, potentially by orders of magnitude.
In this case MBH accretion
is independent of local ISM density and of the velocity of the MBH
relative to the host halo, as the `fuel supply' for accretion
travels with the black hole.

To get a sense of the possible mass-accretion rates allowed by the mechanism,
we need to estimate the gas density in the residual baryonic core. We will
assume that all the baryons in the original system cool and collapse down
to a radius of order $r_{b} \sim 0.1 r_{vir}$ (where $r_{vir}$ is the virial
radius of the original system), at which point thermal pressure or angular
momentum halts the collapse (equivalently we could assume that only a fraction
of the baryons collapse, but that $r_{b}$ is slightly smaller).
If all baryons are in the form of gas and the baryon fraction in
the satellite is cosmological, i.e. $M_{b} = (\Omega_b/\Omega_m)
~M_{vir}$, then the mean gas density is
\begin{equation}
\rho_g =  \frac{3 M_{b}}{4 \pi r_{b}^3} =\frac{1}{0.001} ~\frac{\Omega_b}{\Omega_m} ~\frac{3 M_{vir}}{4 \pi
r_{vir}^3} \approx
 2.39 \times 10^{2} \frac{\Omega_b}{\Omega_m}~\frac{M_{vir}}{r_{vir}^3}
\end{equation}
Substituting this into equation (\ref{eq:Lbol}) and using $c_s \approx 10 {\rm ~km ~s}^{-1}$ for ISM at $10^4$
Kelvin we can determine the mass-accretion rate
\begin{eqnarray}
    \dot{m} & \approx & 6.25 \times 10^{-8} \left(\frac{M_{MBH}}{M_{\odot}}\right)
\frac{\Omega_b}{\Omega_m} \left(\frac{M_{vir}}{10^5
M_{\odot}}\right) \nonumber \\
& & \times \left(\frac{r_{vir}}{{\rm kpc}}\right)^{-3} \left(\frac{c_s}{10 {\rm ~km ~s^{-1}}}\right)^{-3}
\end{eqnarray}
and thus the luminosity. Here $\dot{m}$ is the mass-accretion rate in units of the Eddington accretion rate
\begin{equation} \label{eq:dimless_accrate}
  \dot{m} \equiv \frac{\dot{M} c^2}{L_E} = 1.53
  \left(\frac{\dot{M}}{10^{17} {\rm g ~s}^{-1}}\right) \left(\frac{M_{\bh}}{M_{\odot}}\right)^{-1}
\end{equation}

\subsection{Accretion mechanisms}
As material gets accreted onto a Black hole, a fraction of its gravitational
potential energy released as radiation. The amount and spectrum of energy
released in such a way depends on the accretion model, that is a
prescription for the physical processes that facilitate the
dissipation of energy as material moves towards the accreting MBH.
We consider two of these.

\subsubsection{Thin disk accretion}
Shakura \& Sunyaev (1973) \nocite{shakura73} proposed a model in which
black holes (BHs) accrete from a geometrically thin but optically thick accretion
disk,  which is considered to achieve the highest radiative
efficiencies, typically with $\eta_{td} \sim 0.1$.
Because the disk is optically thick, radiation is emitted
locally with a thermal spectrum and black body
temperature
\begin{equation} \label{eq:thindisktemperature}
    T(r) = \left(\frac{3 G M_{\bh} \dot{M}}{8 \pi r^3 \sigma}\right)^{1/4}
\end{equation}
where $\dot{M}$ and $r$ are the mass-accretion rate and radial
distance from the MBH respectively. We can convert these into
dimensionless quantities and express the accretion rate in units of
the Eddington accretion rate equation (cf.~equation \ref{eq:dimless_accrate}) and
the radial distance from the MBH in units of the MBH
Schwarzschild radius, $r_{\bh}$
\begin{equation}
 \r \equiv \frac{r}{r_{\bh}} = \frac{r ~c^2}{2 ~G ~M_{\bh}} =
0.339  \left(\frac{r}{{\rm km}} \right) \left(\frac{M_{\bh}}{M_{\odot}}\right)^{-1}
\end{equation}

The corresponding spectrum integrated over the whole disk is given by
\cite{frank85}
\begin{equation} \label{eq:td_spectrum}
    \frac{\d L_{\nu}}{\d \nu} = A_0 \frac{h \nu^3}{c^2}
\int^{r_{td}}_{r_{lso}} \frac{\r {\rm d}\r}{e^{h \nu/k T(\r)} - 1}
\end{equation}
where $r_{lso} = 3 ~r_{\bh}$ is the last stable orbit radius of the MBH and $A_0 =
4 ~\pi$ is the area of a spherical shell at unit distance from the
MBH. However, in what follows we treat $A_0$ more generally as a normalisation constant.
All our computations were carried out with $r_{td} = 1000 r_{\bh}$ and
$A_0$ was consequently renormalised by setting $\int (\d L_{\nu}/\d \nu) {\rm
  d}\nu = L_{bol}$.
The resulting spectra are shown in the panels on the right in figure
\ref{fig:accretionspectra}.

In general the accretion flow near the BH may become optically thin and
absorption of photons becomes less important than scattering, which
results in an increase of the mean energy of the photons radiated
away. The emergent flux is thus non-thermal and has a higher
temperature. Consequently in the inner parts of the disk near the MBH,
equations \ref{eq:thindisktemperature}
and \ref{eq:td_spectrum} are not strictly valid anymore. Another
prescription to account for the higher frequency photons is
needed.

This high-frequency emission also has important observational consequences.
MBHs
are only clearly identified as BHs via an X-ray signature. However,
from  equation \ref{eq:thindisktemperature} the peak frequency of the
photons from a standard  thin disk scales as $\nu_{max} \propto T(r_{lso}) \propto
M_{\bh}^{1/4}$. For the MBHs any resultant flux in the X-ray range is
only significant at energies below about 1 keV. A standard thin disk
alone would therefore not provide the X-ray flux by which to identify MBHs.

One could tackle this problem phenomenologically by adding a power-law
spectrum extending into the X-ray range. A power law X-ray spectrum is
well established for active galactic nuclei (AGN), which are driven by SMBHs, and is also found
in recent observations of ultra-luminous X-ray sources in nearby galaxies,
which could well be accreting MBHs \cite{colbert99,wang02}.
However, instead we consider a different model that has recently been
developed particularly with a view to explaining the emission spectra
from accreting SMBH in low-luminosity AGN. This is an advection
dominated accretion flow.

\subsubsection{Advection dominated accretion flows (ADAFs)}
In the ADAF model, once the accretion rate drops
below some critical rate
$\dot{m}_{crit}$, which we define below,  accretion
switches to an `advection dominated accretion flow' (ADAF)
\cite{ichimaru77,rees82,narayan94,manmoto97}.
Because of the low accretion rates the density of the accreted gas is
low, too. The gas, not being able to cool, efficiently stores thermal
energy instead of radiating it away and carries it into the BH.
On the other hand, because the accretion flow is optically thin the
spectrum of the emergent radiation is modified by scattering
processes. The main processes are synchrotron emission, bremsstrahlung
and comptonisation. For more details the reader is referred to the
review by  Narayan, Mahadevan \& Quataert (1998) \nocite{narayan98} and references therein.

The spectrum emitted from a BH accreting in this way depends on a
range of parameters such as the mass-accretion rate , the viscosity
parameter and the relative contribution of magnetic fields.
Typically the spectrum has to be determined numerically, but can be
approximated by analytic scaling relations \cite{mahadevan97}.
Using these, the total radiative efficiency for an ADAF, $\eta_{adaf}$ can be
expressed in terms of that of a thin disk, $\eta_{td}$ as
\begin{equation}
\eta_{adaf} = 0.2 \eta_{td} \left(\frac{\dot{m}}{\alpha^2}\right)
g(\theta_e)
\end{equation}
for a typical set of parameters as given in Mahadevan (1997). Here $\dot{m}$ is the mass-accretion rate in terms of the Eddington
accretion rate, $\alpha$ is the viscosity parameter ($\alpha \sim 0.1$
to $0.5$), $\theta_e = k_B ~T_e/(m_e ~c^2)$ is the electron kinetic
energy in units of their
rest mass energy. $g(\theta_e) \approx 11.5 (T_e/10^9 {\rm K})^{-1.3}$
is a measure of whether electrons in the flow move at
relativistic speeds or not. Typically $g \sim 1$ to $10$ for
electron temperatures $2\times 10^9 < T_e < 10^{10}$ K.
We can determine an equilibrium electron temperature which is
approximately
\begin{equation}
  \left(\frac{T_e}{10^9 {\rm K}}\right) \approx
  \left(\frac{\alpha}{0.3}\right)^{-0.06}
  \left(\frac{r_{min}}{3}\right)^{0.1} ~M_{\bh}^{0.06}
  ~\dot{m}^{-0.04}
\end{equation}

Generally the effect of ADAFs is to reduce the overall radiative
efficiency compared to that from thin disk accretion, although
individual regions of the accompanying spectrum, especially the X-ray
band, tend to be much stronger than for a thin disk.

As mentioned above the thin disk and ADAF regime can be delineated by the critical mass
accretion rate $\dot{m}_{crit}$ which is given by
\begin{equation}\label{eq:m_crit_ADAF}
 \dot{m}_{crit} \ge 0.28 \alpha^2
\end{equation}
For rates above $\dot{m}_{crit}$ the ADAF mechanism breaks down and in this regime we will
work with the thin disk model.
For low values of $\alpha \leq 0.1$  accretion may switch to a `convection
dominated accretion flow' (CDAF) \cite{ball01}, which is characterised
by much stronger suppression of mass accretion. We will neglect this
possibility here as we work with $\alpha = 0.3$ in the
computation of spectra and luminosity functions for our model.
This is also the fiducial value used in the work of Mahadevan
(1997). We will briefly consider changes in $\alpha$ below.

In paper I we have seen that MBHs do not accrete at
more than 1/10 of the Eddington rate with most accreting at less than
1/100. Thus the ADAF model can be applied to most MBHs we are dealing
with.

The three dominant contributions to the continuum spectrum of an ADAF are those
from synchrotron and bremsstrahlung emission as well as
comptonisation, which can all be separately approximated by scaling
laws \cite{mahadevan97}.
In the radio-submillimeter range synchrotron emission dominates and the
luminosity (per unit frequency) scales as
\begin{eqnarray}\label{eq:adaf_synchro_scaling}
 \frac{\d  L_{sync}}{\d \nu} & \simeq& 3.51 \times 10^{13} ~
  \left(\frac{\alpha}{0.3}\right)^{-4/5} \left(\frac{T_e}{10^9 ~{\rm
  K}}\right)^{2/5} \nonumber \\
  & & \times \left(\frac{M_{\bh}}{M_{\odot}}\right)^{6/5}
  ~\dot{m}^{6/5} ~\nu^{2/5} {\rm ~erg
  ~s}^{-1} ~{\rm Hz}^{-1}
\end{eqnarray}
up to a peak frequency $\nu_{p}$ given by
\begin{eqnarray}
  \nu_p & =& 1.71 \times 10^{17} \left(\frac{\alpha}{0.3}\right)^{-1/2}
  \left(\frac{T_e}{10^9 {\rm K}}\right)^2
  \left(\frac{M_{\bh}}{M_{\odot}}\right)^{-1/2} \nonumber \\
  & & \times \left(\frac{r_{min}}{3}\right)^{-5/4} ~\dot{m}^{3/4} {\rm Hz}
\end{eqnarray}
Photons from synchrotron emission can get comptonised, which raises
their energy up to a limit given by the electron temperature $k_B ~T_e$.
Since most synchrotron photons are emitted at $\nu = \nu_p$ we can
make the simplifying assumption that this is the initial frequency for
all photons to be comptonised.
The comptonisation spectrum is then simply
\begin{equation} \label{eq:adaf_compton_scaling}
  \frac{\d L_{compt}}{\d \nu} \simeq \frac{\d L_{sync}}{\d \nu}(\nu_p) \left(
  \frac{\nu}{\nu_p}\right)^{-\alpha_c}
\end{equation}
where the power law index is primarily a function of electron temperature
and mass-accretion rate
\begin{equation} \label{eq:compton_power_index}
  \alpha_c \equiv - \frac{\ln \left[23.87 \dot{m}
  \left(\frac{\alpha}{0.3}\right)^{-1}
  \left(\frac{r_{min}}{3}\right)^{-1/2} \right]}{\ln \left[1 + 4
  ~\theta_e +
  16 \theta_e^2 \right]}
\end{equation}

\begin{figure*}
  \includegraphics[width=13cm]{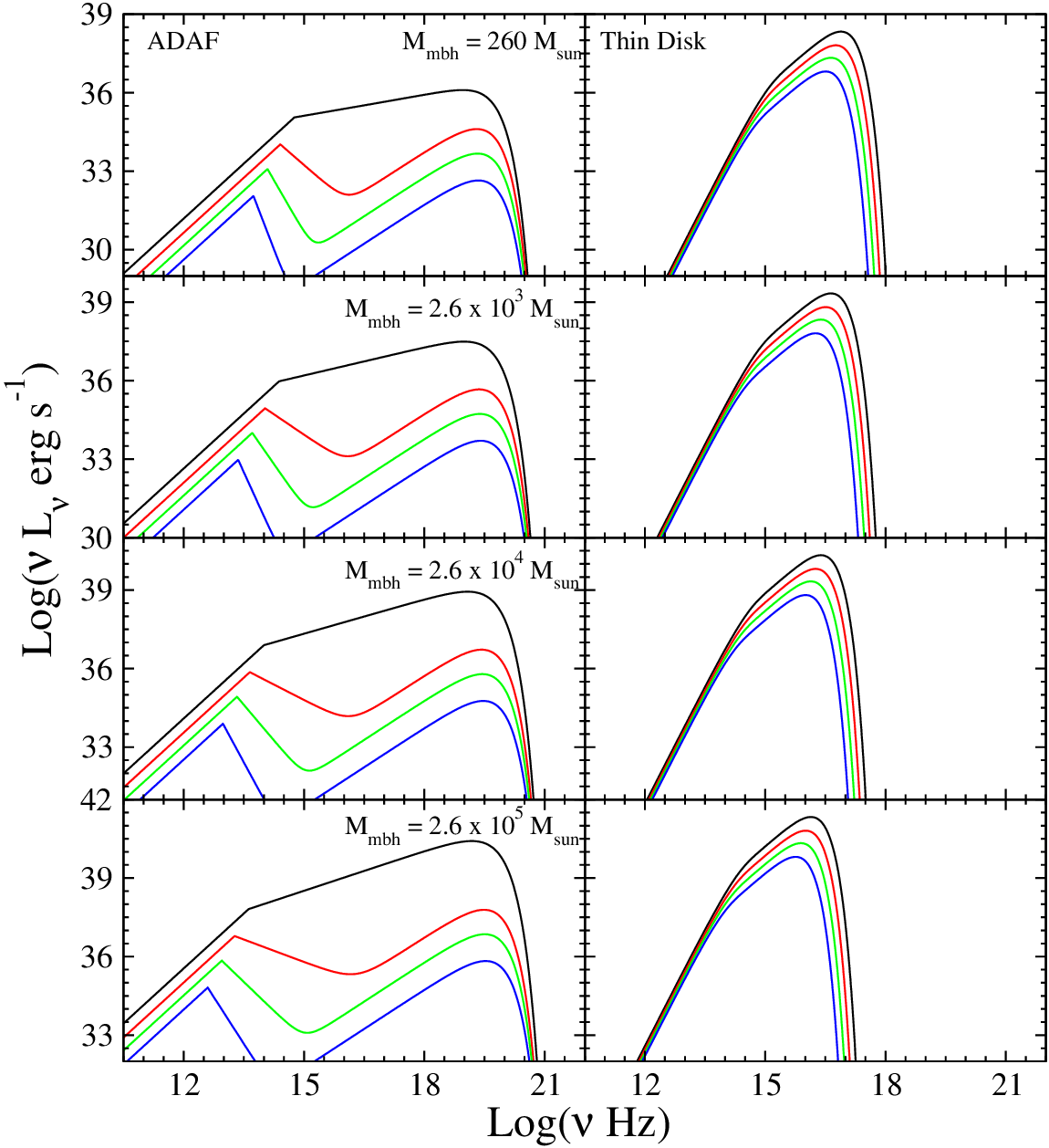}
  \caption{Emission spectra from ADAF (left) and thin disk (right)
  accreting MBHs in the mass range between $2.6\times10^2$ to
  $2.6\times10^5$ \Msun. In each panel the curves from top to bottom
  correspond to mass-accretion rates $\dot{m} = \{1, 0.3, 0.1, 0.03\}
  ~m_{crit}$, where $m_{crit} = 0.0252$ for $\alpha = 0.3$.}
  \label{fig:accretionspectra}
\end{figure*}

There is a constant contribution from bremsstrahlung emission that
also tails off exponentially at the maximum frequency implied by the
electron temperature. Bremsstrahlung only becomes prominent in the
UV/soft X-ray range if $\alpha_c > 1$, such that power from Compton
emission effectively decreases with increasing frequency
\begin{eqnarray}\label{eq:adaf_brems_scaling}
  \frac{\d L_{brems}}{\d \nu} & \simeq& 1.02 \times 10^{17}
  \left(\frac{\alpha}{0.3}\right)^{-2}
  \ln\left[\frac{r_{max}}{r_{min}}\right] \nonumber \\
  & & \times F(\theta_e) \left(\frac{T_e}{10^9 {\rm K}}\right)^{-1} e^{- h \nu/k_{\small B}
  T_e} \nonumber \\
  & & \times \left(\frac{M_{\bh}}{M_{\odot}}\right) \dot{m}^2 {\rm erg ~s^{-1} ~Hz^{-1}}
\end{eqnarray}
If $\theta_e < 1$, $F(\theta)$ takes the form
\begin{eqnarray}
  F(\theta_e) &=& 4\left(\frac{2\theta_e}{\pi^3}\right)^{1/2} (1 + 1.78
    \theta_e^{1.34}) \nonumber \\
    & & + 1.73 \theta_e^{3/2}(1 + 1.1\theta_e +
    \theta_e^2 - 1.25 \theta_e^{5/2})
\end{eqnarray}
whereas for the case $\theta_e > 1$,
\begin{eqnarray}
  F(\theta_e) &=& \left(\frac{9 \theta_e}{2 \pi}\right) [\ln(1.123 \theta_e + 0.48) +
    1.5] \nonumber \\
  & & + 2.3 \theta_e (\ln1.123 \theta_e + 1.28)
\end{eqnarray}
The resulting spectrum is shown in the left panels of figure \ref{fig:accretionspectra}
for a range of MBH masses and accretion rates ranging from $\dot{m} =
\dot{m}_{crit}$ to $\dot{m} = 0.03 ~\dot{m}_{crit}$, where
$\dot{m}_{crit} = 0.0252$ for $\alpha = 0.3$ which we use for all our
computations.

We see that only ADAF accreting MBHs display significant X-ray
emission. It is therefore mainly MBHs that could be identified
as such by their X-ray signature.
Conversely, while optical and UV observations may be used as
follow-ups, the sources that are most luminous in this range are
accreting from a thin disk and so will not have any associated strong
X-ray emission. \\
\vspace{3mm}\\
{\it X-ray hardness ratios}\\
For our predictions we have computed the X-ray emissions for the energy range between $0.5
< h~\nu < 2.0$ keV and $2.0 < h~\nu < 10.0$ keV
, which we call the soft and hard X-ray bands respectively.
Both bands lie in the Bremsstrahlung part of
the ADAF spectrum where $\d L/\d \nu = 1$, i.e. the photon counts
from both bands are the same; their {\it hardness ratio} is equal
to 1. Unfortunately, observations from X-ray binaries (XRBs)
can yield similar values, so it will not be easy to distinguish
between MBHs and XRBs on the basis of this alone.
The combined thin disk + ADAF model that we used, means that we can
also have sources with large soft X-ray luminosities and very small
hardness ratios. This is the case particularly for seed MBHs for which
thin disk accretion reaches furthest into the soft X-ray regime, and
there actually dominates over the respective ADAF contribution
(cf.~figure \ref{fig:accretionspectra}). In section \ref{sec:xray_em}
we see that this leads to a distinctive `bump' in the soft band
luminosity function.
These sources should be identifiable by their very low hardness ratio.

\section{Predicted Luminosity Functions}
\label{sec:lum_functions}
\begin{figure}
  \label{fig:bol_lum}
  \includegraphics[width=8cm]{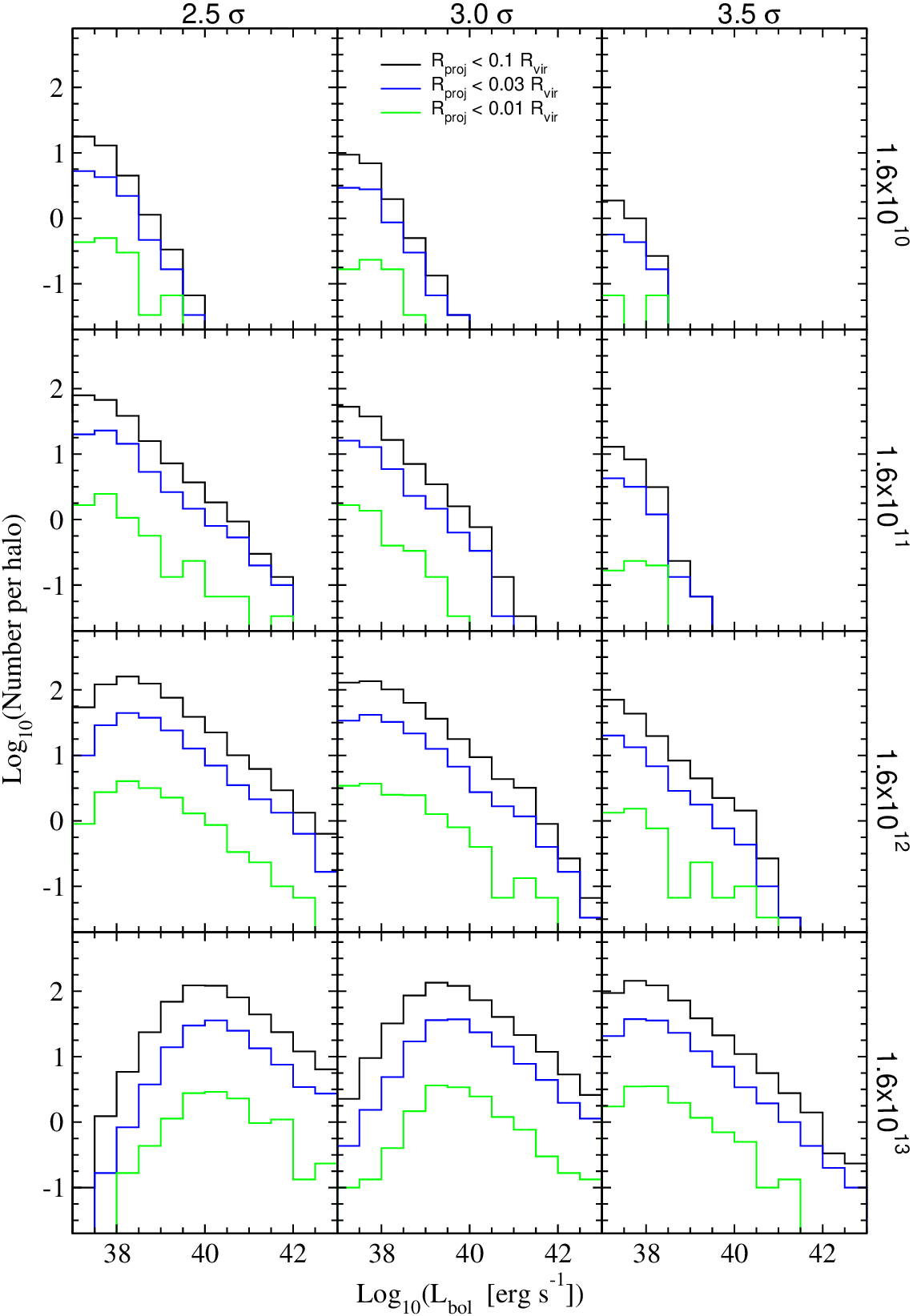}
  \caption{Bolometric luminosity of MBHs accreting from baryonic
    cores for models C, A and D (left to right panels). Shown are the
    sources whose line of sight falls within some
    projected distance from the host centre.}
  \label{fig:Lbol_project}
\end{figure}

Using the spectral models outlined above we have computed optical and
X-ray luminosities of accreting MBHs in present-day haloes.
For comparison we show in figure \ref{fig:bol_lum} the bolometric luminosity functions, obtained
in paper I.

\subsection{Optical emission} \label{sec:optical_em}

\begin{figure}
  \includegraphics[width=8cm]{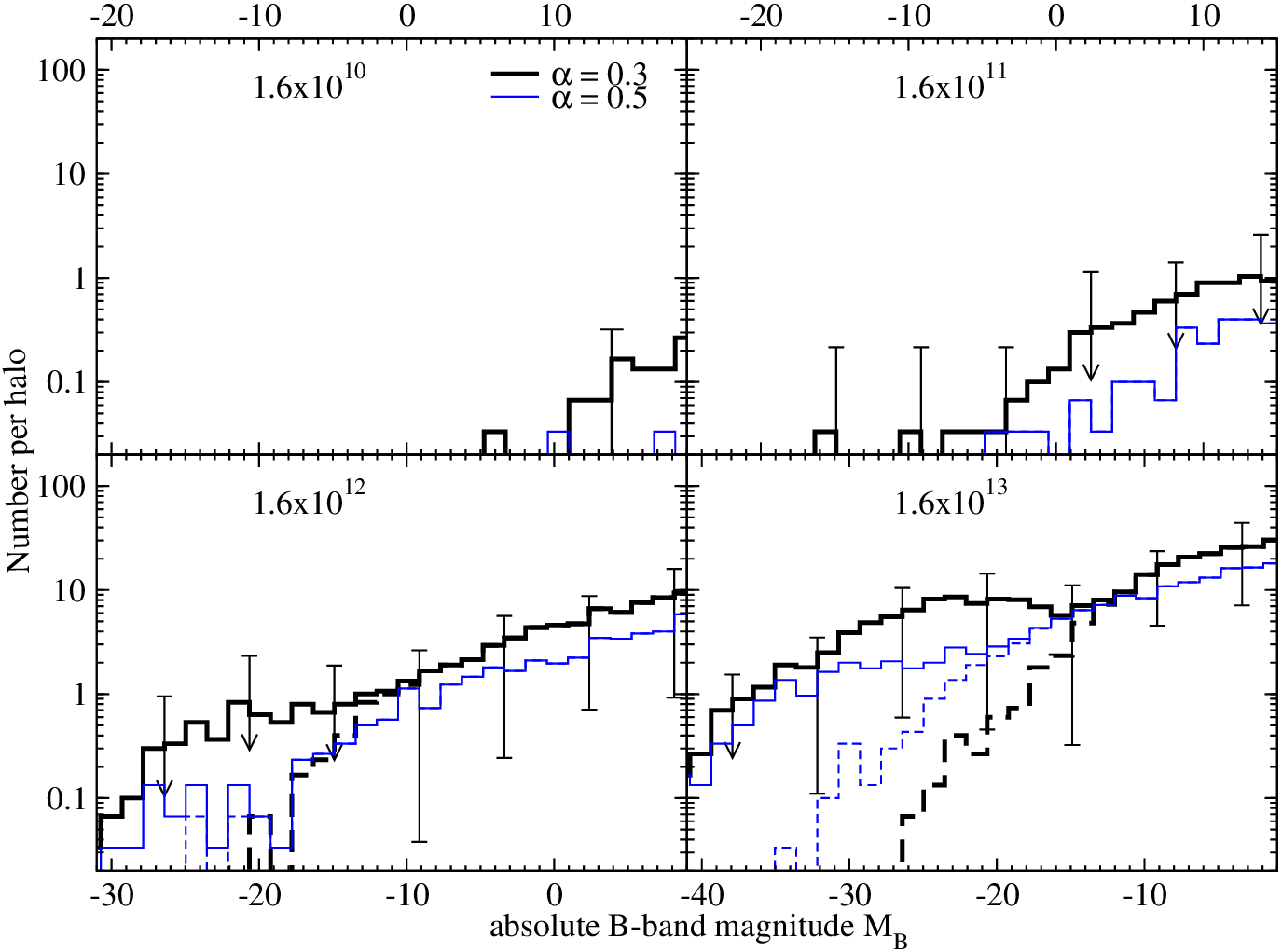}
  \caption{Average B-band
    luminosity functions for MBHs in haloes of various masses,
    for two different values of $\alpha$ (two sets of lines). The
    dashed lines show the contribution from ADAF accretion. Thin disk
    emission dominates at large luminosities.
  }
  \label{fig:Bmag}
\end{figure}
For the MBH masses we are considering here, the optical region lies
within the Rayleigh-Jeans part of the extended black body spectrum
that characterises the corresponding thin disk emission.
In this part, $k_B T(r_{td}) \ll h\nu \ll k_B T(r_{lso})$ where the
spectrum can be approximated by a power law
\begin{equation}
    \frac{\d L}{\d \nu} = A_1 \left(\frac{\nu}{{\rm Hz}}\right)^{1/3}
\end{equation}
and the constant $A_1$ depends on the MBH mass and accretion rate as
\begin{eqnarray}
    A_1  &=& 2.57 ~A_0 ~h^{-5/3} ~k_B^{8/3} ~c^{-2} ~T(r_{lso})^{8/3}
    \\
    &\approx& 1.64 \times 10^{11} ~A_0 ~\dot{m}^{2/3}
    ~M_{\bh}^{4/3} ~{\rm erg ~s^{-1} ~Hz^{-1}} \nonumber
\end{eqnarray}

In the ADAF regime it is primarily synchrotron emission and
comptonisation that contribute to the emission in the optical/UV
range, with the corresponding scalings given in equations
\ref{eq:adaf_synchro_scaling} and \ref{eq:adaf_compton_scaling}.

Figure \ref{fig:Bmag} shows the B-band luminosity function. The dashed
line represents the contribution from ADAF emission with the rest,
primarily at large luminosities, coming from thin disk
emission. Apart from our fiducial value for the viscosity parameter,
$\alpha = 0.3$ (thick lines), we have also given results
  for $\alpha = 0.5$ (thin lines). While the results for ADAF emission are 
typically not too
sensitive to changes in ADAF parameters, results are affected
more strongly by changes in $\alpha$, but only through its direct impact on
the value of $\dot{m}_{crit}$ (cf.~equation \ref{eq:m_crit_ADAF}).
As a result the most important effect of increasing $\alpha$ is to extend the
number of ADAF emitting sources to larger luminosities. Since a thin
disk radiates more efficiently than an ADAF this results in a
lower abundance of sources at very large magnitudes in the optical
range. For smaller values of $\alpha$ we get a correspondingly larger
number of thin disk accreting sources. However, we found that for a
value of $\alpha = 0.1$, for instance, the resulting luminosity
function in the B-band displays a significant `dip' which is also present 
to some degree for $\alpha = 0.3$ in the case of the most massive
$1.6\times10^{13}$ \Msun halo in figure \ref{fig:Bmag}.
We will deal with the results for the X-ray range in the next section,
but note at this point that a higher (lower) value of
$\alpha$ would result in a larger (lower) abundance of sources in the
X-ray regime, emission in which is mainly produced in ADAFs.

\begin{figure*}
  \vbox{
    \includegraphics[width=8.5cm]{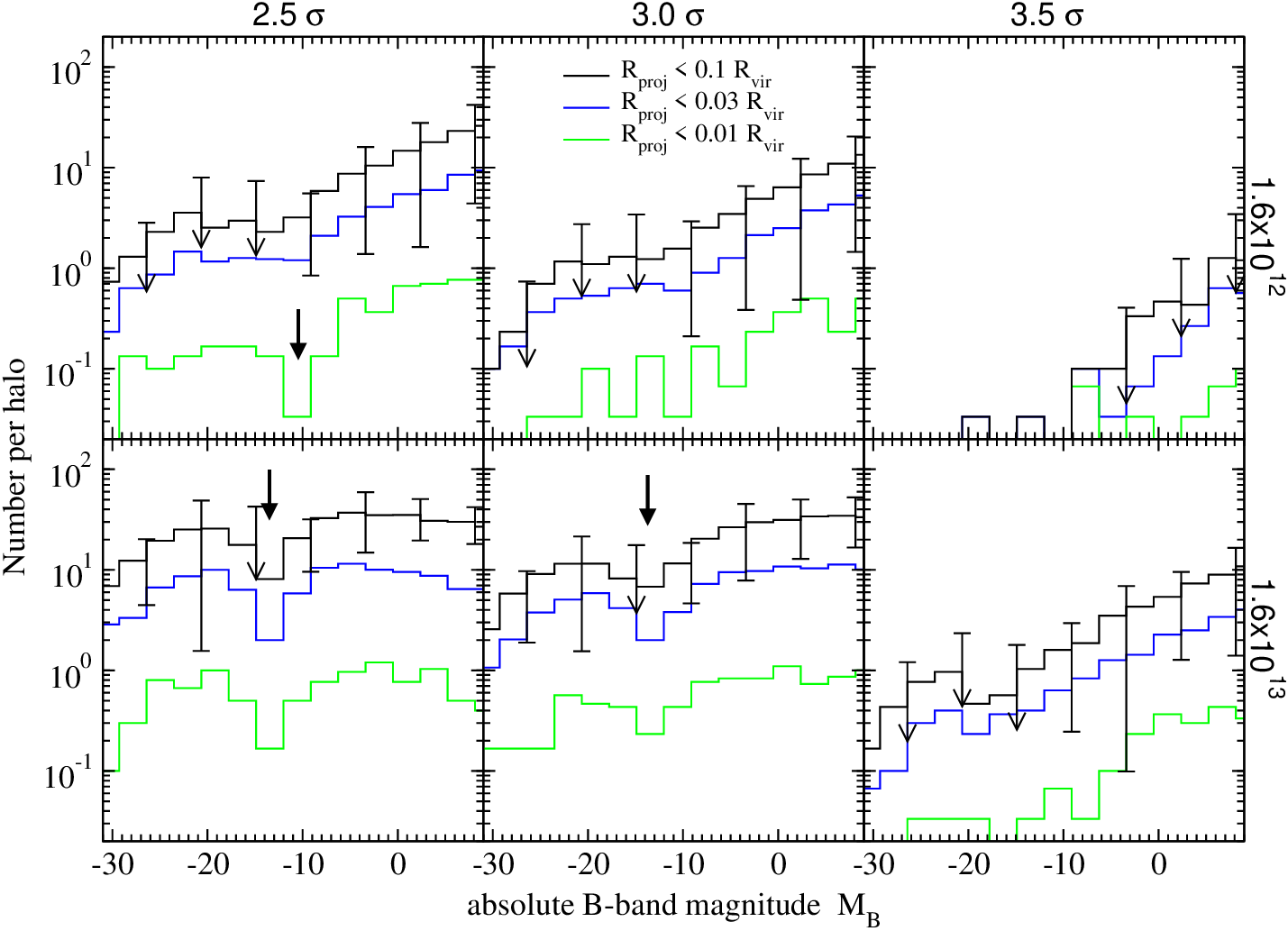}
    \includegraphics[width=8.5cm]{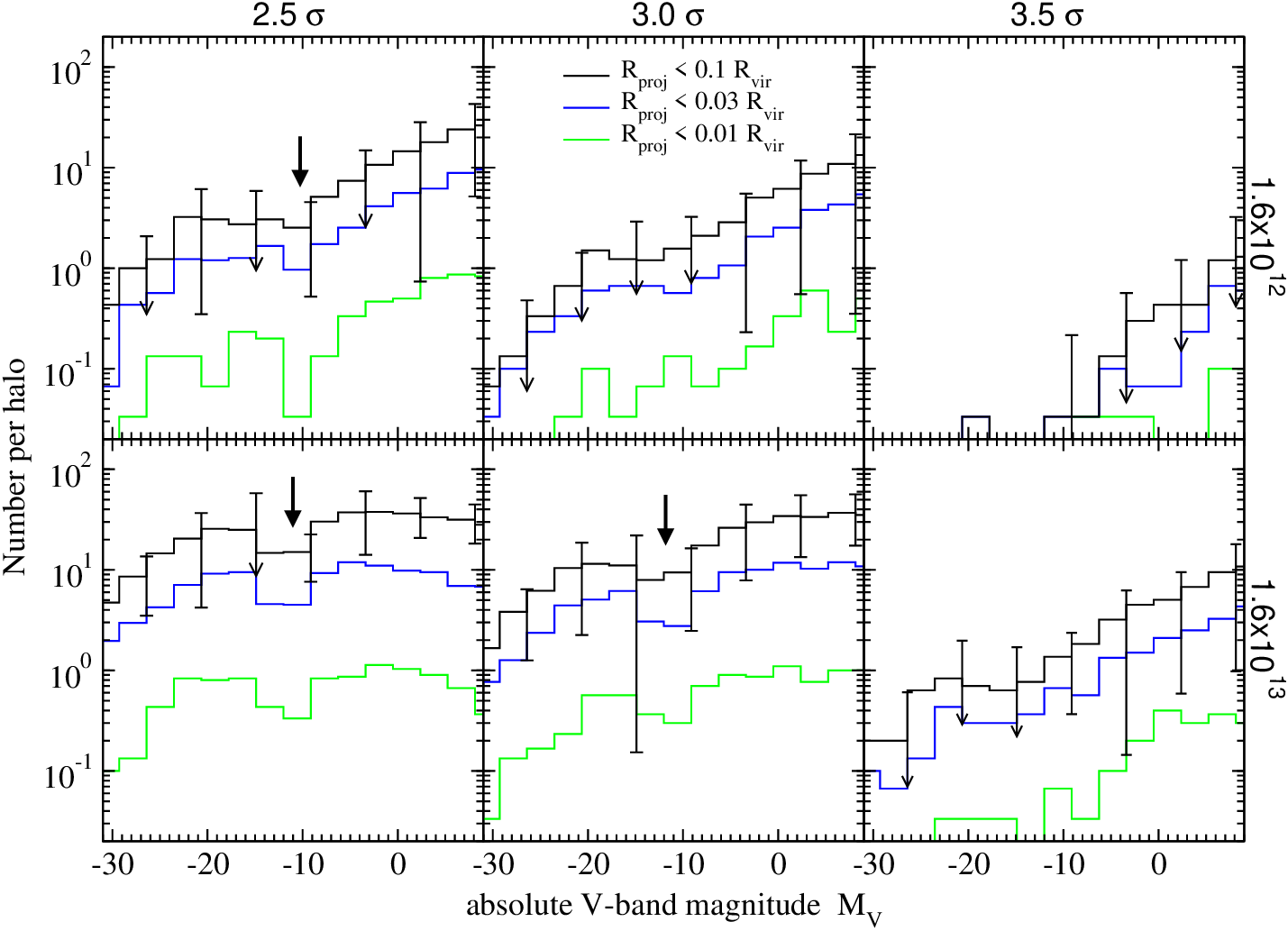}
  }
  \caption{Optical luminosity function for MBHs
    accreting from baryonic
    cores within various projected distances from the host centre.
    Results are shown for models C, A and D (three sets of panels,
    left to right)
    in the B and V-band (left-hand and right-hand figures). The bold
    arrows delineate where luminosities are dominated by thin disk
    emission and ADAF emission respectively.
  }
  \label{fig:optical_project}
\end{figure*}

In figure \ref{fig:optical_project} we show the B and V band luminosity function of MBHs
within some projected distance from the host centre normal to the line of sight. Here we
have included all MBHs within the respective host virial radius along the lines of
sight passing within some projected distance from the centre. We will
subsequently refer to this as the {\it projected} luminosity
function. The projected luminosity function can more easily be compared
with observations, which typically refer to sources within galaxies.
In each case the function is plotted for MBHs sources
within $R_{proj} = 0.1, 0.03, 0.01 ~R_{vir}$ from the host centre, where the latter
corresponds roughly to the outer radial extent of the galaxies in the
haloes. Across all halo masses we find that a fraction of about 30, 9 and 0.9 \%
respectively of all halo MBHs are located within the projected
distances listed above. Since there is no discernible difference in
the relative distribution of MBH masses at different radii
(paper I), this means that for any one MBH in a given
mass, accretion rate or luminosity range and within $R_{proj} = 0.1
R_{vir}$, there are about two more in the halo.
We have marked with
bold black arrows the luminosity range below which ADAF emission
dominates.

It is important to note that these luminosity functions only provide
an upper limit on what could actually be detected. We have here neglected
the effect of absorption which is expected to significantly reduce the
flux of optical and UV photons due to dust and neutral hydrogen along 
the line of sight. The
latter affects particularly the light from sources whose line of sight
passes closely to the galactic centre, such as those with $R_{proj} <
0.01 R_{vir}$ and located on the far side of the halo.
We would expect this effect to be less important in gas-poor
ellipticals.

Even if absorption is accounted for, the
large luminosity of the accreting MBHs by itself may not be sufficient to
identify them as such. As they appear as point sources, they could
also be XRBs or background AGN.
For the largest accretion rates, $\dot{m} \sim 0.1$ (paper I), at which
MBHs accrete from a thin disk
in our model, it is possible
that emission occurs in the form of (mini) jets which may be
observable even though the optical and UV output from across the
accretion disk may not.

\subsection{X-rays} \label{sec:xray_em}
\begin{figure*}
  \vbox{
    \includegraphics[width=8.5cm]{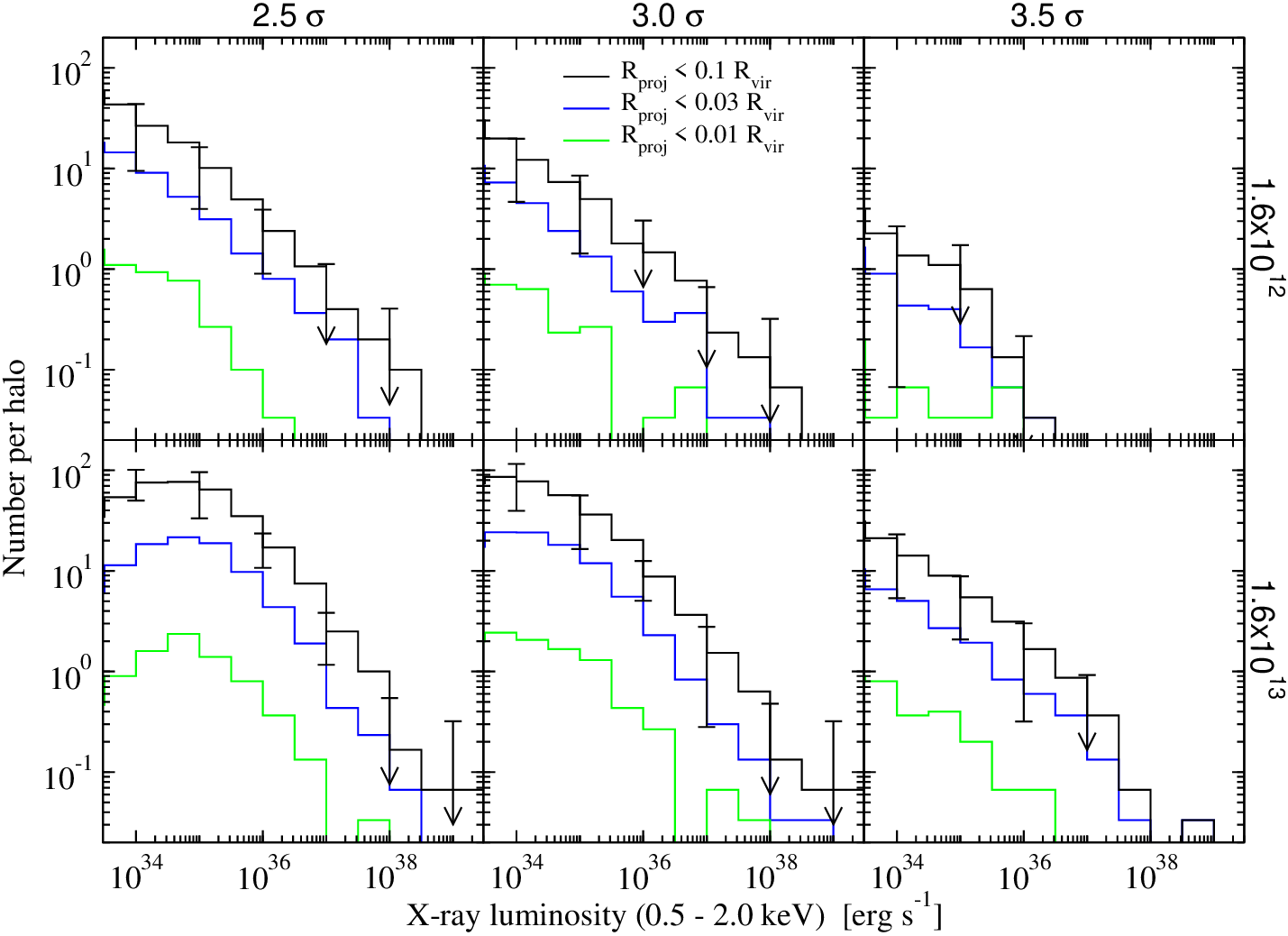}
    \includegraphics[width=8.5cm]{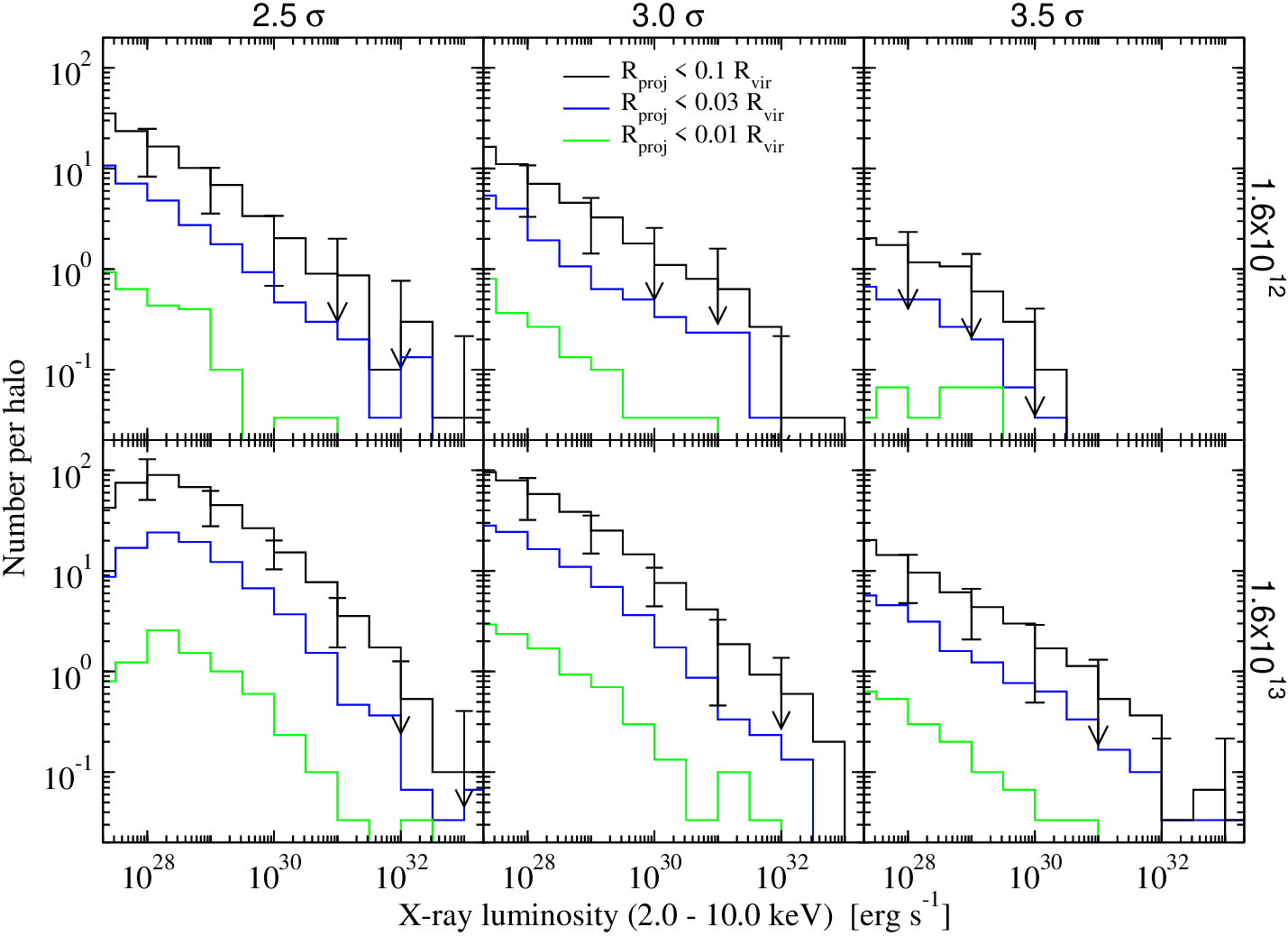}
  }
  \caption{X-ray luminosity function of MBHs with
    (line-of-sight within) various projected distances from the host
    centre (see also text).
    Results are shown for models C, A and D (three sets of panels,
    left to right)
    and X-rays in the energy range from 0.5 - 2.0 keV (left-hand figure) and
    2.0 - 10.0 keV (right-hand figure).}
  \label{fig:Xray_project}
\end{figure*}
In our model X-rays are primarily emitted by ADAF accreting MBHs.
Emission in the X-ray range is dominated by comptonisation or
bremsstrahlung depending on the Compton spectral power law index, $\alpha_c$
(eq. \ref{eq:compton_power_index}). The two contributions are
determined by the scaling relations equations \ref{eq:adaf_compton_scaling} and
\ref{eq:adaf_brems_scaling}.
In figure \ref{fig:Xray_project}  we present the projected X-ray luminosity
function, both in the 0.5 - 2.0 keV and 2.0 - 10.0 keV energy range.
Again we have ignored the issue of absorption here, which can be important
primarily for the softer 0.5 - 2.0 keV X-ray band.

X-ray emissions are the primary signature to look for in
observations, while optical and UV detections
are probably useful only for follow-up, due to the range of potential sources
and the problem of absorption. Another secondary diagnostic would be to see
whether any correlations exist between different parts of the emitted spectrum.
However, this would be extremely dependent on the accretion
model. For instance, in the ADAF model that we used, the shape of the spectrum near
the exponential cut-off in the X-ray regime and in the
radio/submillimeter range is essentially unaffected by the mass and
accretion rate of the MBH (cf.~figure \ref{fig:accretionspectra}). If a
MBH accretes via an ADAF we would therefore expect a correlation of
the radio and X-ray luminosities.

\section{Comparison with Observations}
\label{sec:observations}
We now compare our predictions with observations. As we have already
noted, X-ray emissions will be the most important trace to look for. We will
therefore confine our comparison to X-ray observational data only and
further more to sources which have been identified as belonging to a galaxy.

In our predictions we have generally not taken into account absorption of MBH emission
either in the originating galaxy as well as ours. To minimise the
former one would want to look for sources in external galaxies that
we see face-on. This is especially important for gas-rich spirals.
To account for the extinction from the IGM and the Milky-Way's ISM
all-sky extinction maps can be used \cite{schlegel98}. These
allow a determination of the average hydrogen column density between
us and the target galaxy.

\subsection{Observations of ultra-luminous X-ray sources (ULXs)}

For the baryonic core densities and accretion mechanisms considered above,
our model predicts the presence of ultra-luminous X-ray sources (ULXs) in
galaxies and their haloes. By ULX we refer to any compact X-ray source
with $L_X \grtsim 10^{38} {\rm erg ~s}^{-1}$. This cutoff is interesting
because it corresponds roughly to the Eddington limit for isotropic
accretion onto stellar-mass black holes. Early observations revealed
that some galaxies contain these very bright sources  outside their
nucleus \cite{fabbiano89}. Subsequently, a number of other studies
discovered ULXs in data from the ROSAT and ASCA satellite missions
\cite{colbert99,makishima00,roberts00,colbert02}. With the advent of
the {\it Chandra} and {\it XMM} satellite missions, the
number of ULX detections has soared. The superior resolution and
sensitivity of these satellites has greatly increased the distance
at which ULXs can be detected, and thus the number of galaxies in which
their properties can be studied.

While `normal' stellar remnants may easily account for the fainter ULXs,
sources which emit more than a few times $10^{39} {\rm erg ~s}^{-1}$
require black holes more massive than those produced by conventional
stellar populations, assuming the emission is Eddington limited and
occurs isotropically. If
ULXs do not emit isotropically, however, but produce beamed emission
that happens to be aligned with the line-of-sight, then
stellar mass BHs can be the source \cite{king01,zezas02}.
A departure from isotropy implies that the Eddington luminosity
is no longer a limiting factor, either, as radiation pressure need not
act along the same axis as the accretion flow.
High mass X-ray binaries (HMXRBs) have been suggested as candidates
for anisotropic ULXs. Within galaxies, ULXs appear to be found predominantly
in star-forming regions \cite{kilgard02,zezas02}, i.e. they seem to be
associated with starburst galaxies or the spiral arms of disk galaxies.
For a number of these galaxies it was found that the ratios of ULX number to
massive star formation rates are similar \cite{smith03}.
Thus many ULXs may be young HMXRBs, which would naturally be
found in or close to star-forming regions.

There is some evidence, however, to suggest that not all ULXs fall into
this category. Multi-colour disk (MCD) fits to ULX spectra indicate that
the maximum colour temperature associated with some sources is
significantly lower than would be expected for a stellar mass BH
system. If the maximum temperature arises from emission close to the
inner radius of the accretion disk, and we assume this
to be the BH's last stable orbit, a lower temperature implies a
larger stable orbit and thus a more massive BH \cite{wang02}.
Furthermore, ULXs have been detected in some gas-poor ellipticals
\cite{jeltema03}, and
Colbert \& Ptak (2002) \nocite{colbert02} find that the number of
ULXs per galaxy is actually higher for ellipticals than for
non-ellipticals. Thus the connection between ULXs and star formation
established for spiral galaxies may not be universally valid.
It seems interesting, therefore, to consider alternative explanations
for these sources, such as isotropically emitting MBHs.

If seed MBHs do account for some of the observed ULXs, there are
a few remaining puzzles to work out.
It is not clear whether our model should predict more ULXs around
spirals or ellipticals, and the morphological information in the
model is probably too crude to test this definitively. Since
ellipticals are generally older systems, however, it is not implausible
that they would have a large number of ancient satellites in
orbit around them. A number of ULXs seem to be located in globular clusters
\cite{jeltema03}, suggesting binaries are the corresponding ULX sources.
In the context of our model, however, some of these globular clusters
might be precisely the baryonic cores we have considered previously.

Ultimately, detailed spectral information should allow us to distinguish
between binaries and MBHs as ULX candidates. Reasonable spectral information
is available for some systems, although fits to such data still assume
model spectra which should be verified independently. For the more distant
sources, the detected count rates are so low that spectral fits cannot
be carried out. Instead a {\it hardness} ratio, the ratio of the number
counts in a soft and a neighbouring hard X-ray band, can be used as
a crude diagnostic. Uncertainties in the correction for absorption
in soft X-ray bands complicate the interpretation of these results, however.
Here we will avoid these complications, using spectral information
only to the extent that it allows X-ray band luminosities to be
derived from raw photon counts. We will not use spectral fits
to make any inference about the ULX source object.

In summary, while binaries may explain the majority of ULXs,
especially in star-forming galaxies, MBHs are not ruled out
as a possible source of some of these objects. In what follows,
we will show how the free-floating MBHs predicted by our model
could account some fraction of all ULXs. The exact fraction
is highly model dependent; for the parameters assumed above,
we will show it can be considerable.

\subsection{Comparing Luminosity Functions}

Table \ref{tab:Xray_observations} gives a summary of the observations
of point sources in individual galaxies that
we are considering. Their inferred masses roughly place them in
a category with the $1.6\times10^{12}$ \Msun halo in our
model. Observed sources are considered roughly within the light
radius of the galaxy. Where observations do cover regions
significantly outside this radius it is typically an area to one
side of a galaxy that happens to be still covered by the detector.
Interestingly, a number of ULXs are
detected in this area as well. Estimates of the expected
number of background objects, such as AGN, supernovae (SN), etc. 
show that they can account
for most of the ULXs observed outside the light radius, although the
error margins are large.
A number of sources outside the bulge have projected locations that
appear to be in the disk. In this case background sources cannot
account for the relatively large number of sources that seem to lie in
the disk. It is assumed on this basis that these sources
are actually located in the disk, where their number can be explained
more naturally.
For sources with apparent locations in the disk as well as outside the
light radius our model offers the alternative explanation that some of these
sources could also be located throughout the halo but inside a column
along the line of sight that projects onto the disk.

\begin{table*}
  \caption{Summary of X-ray observations. The first four galaxies are spirals;
    NGC 720 is an elliptical. All observations
    were carried out with the ACIS instrument on {\it Chandra}, except
    for the observation of M31 which was observed with {\it Chandra}'s HRC instrument.}
  \label{tab:Xray_observations}
  \begin{minipage}{140mm}
    \centering
    \begin{tabular}{lllll}\\ \hline\hline
      Galaxy     & Energy range      & Region surveyed & \# ULXs   & Reference \vspace{0.3cm} \\  \hline
      M31      (S)   & 0.1 - 10 keV      & $R_{proj} < 1.13 {\rm kpc}$  &  3            & \cite{kaaret02}\\ \hline
      M81     (S)    & 0.3 - 8  keV      & $R_{proj} > D_{25}$    &  2            & \cite{swartz03}\\
      &                   & disk            &  10            & \\
      &                   & bulge           &  5            & \\
      & 0.2 - 8  keV      & disk ($R_{proj} \lesssim 8.5 {\rm kpc}$) & 5   &\cite{tennant01} \\
      &                   & bulge           &  3            & \\ \hline
      NGC 6964  (S)  & 0.5 - 5  keV      & $R_{proj} \lesssim 13 {\rm kpc}$& 15 & \cite{holt03}\\ \hline
      NGC 1068  (S)  & 0.4 - 5  keV      & $R_{proj} < D_{25}$ & 40        & \cite{smith03}\\ \hline
      NGC 720 (E)& 0.3 - 10 keV      & $R_{proj} \lesssim 25 {\rm kpc} $ & 41 &\cite{jeltema03}\\ \hline\hline
    \end{tabular}
  \end{minipage}
\end{table*}

Given the uncertainties and free parameters in our model, we cannot
place a strong constraint on MBHs from ULX observations alone.
Nonetheless, it is interesting to make a quantitative comparison
between the observations and our model predictions, particularly
since the predictions scale in a straightforward way as the
number or mass of the seeds, or the density of the baryonic cores,
varies. Since our models often predict one or fewer luminous sources
per system, the comparison with observations will be statistical;
this is all the more true since we predict a large halo-to-halo
scatter in the number of MBHs, and thus in the X-ray luminosity
functions. In general, even if MBHs accounted for all ULXs, we
would expect a variation in the luminosity functions of individual
galaxies that was comparable to the halo-to-halo scatter in our results.

\begin{figure*}
  \includegraphics[width=15.5cm]{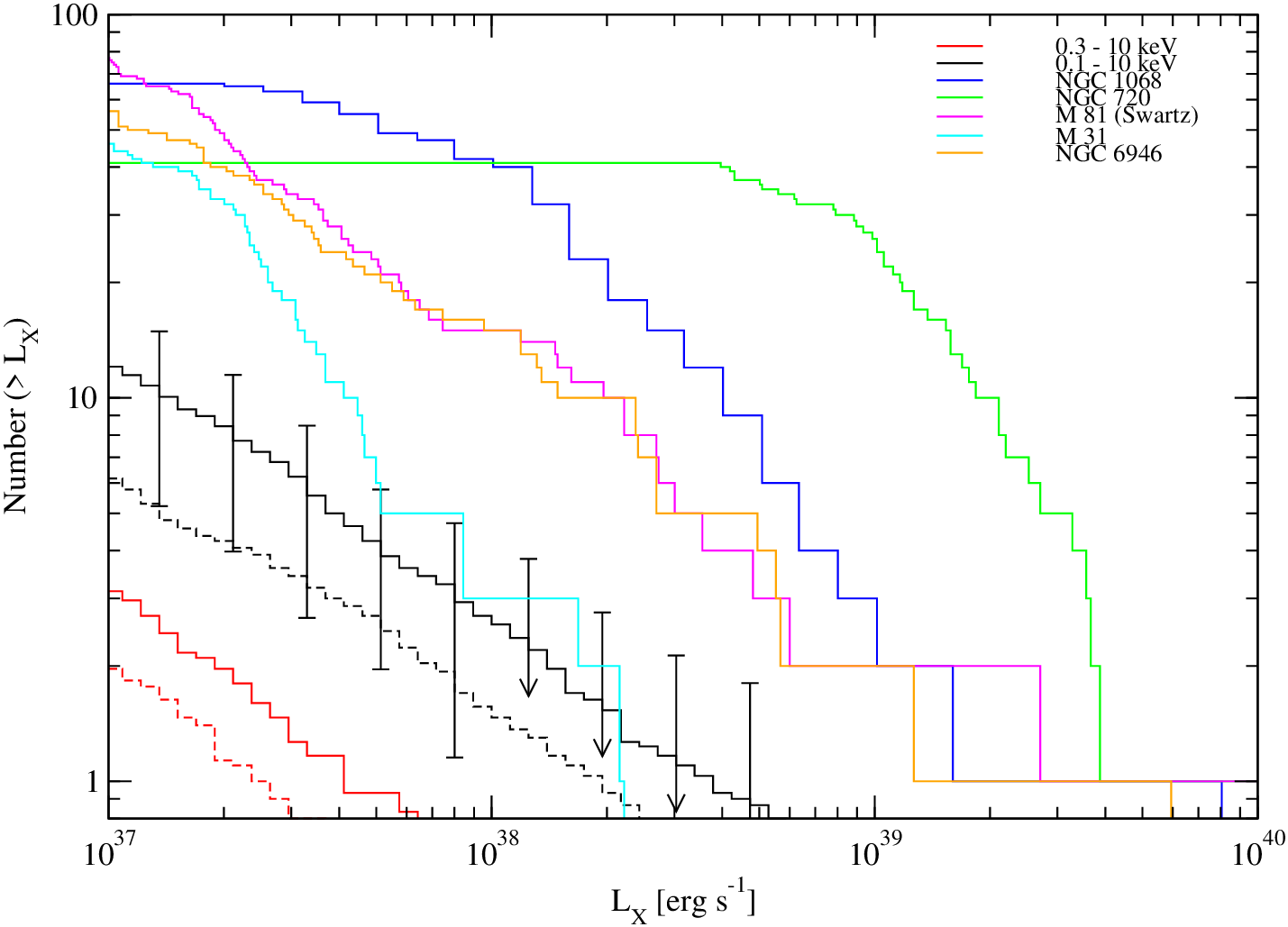}
  \caption{Cumulative X-ray luminosity functions for five galaxies observed 
  by {\it Chandra} (see text), compared with our model predictions.
  The solid black line with error bars shows our predicted cumulative
  luminosity functions for model C in the 0.1-10 keV band, while the
  dashed line immediately below it shows the same for model A.
  The two lines below this are for models C and A in the 0.3-10 keV
  band. In each case only those sources with projected radii $< 30$ kpc
  from the galactic centre are included.
  }
  \label{fig:Xray_observations}
\end{figure*}

We first look at how our results compare with observations of sources
in individual galaxies.
Figure \ref{fig:Xray_observations} shows the cumulative X-ray
luminosity function from observations and those that we predict for a
$1.6\times10^{12}$ \Msun halo, which would be expected to host galaxies
of the mass we are comparing with. For
our predictions we have chosen two different X-ray bands as shown in
the figure to match those from the observations more closely.
Our predictions are consistent with the data from the individual
galaxies if we only consider the ADAF emissions. Inclusion of the thin
disk component leads to a much flatter slope and an over prediction of
sources at the high luminosity end. The flat slope essentially bridges
the gap in the differential luminosity function between ADAF accreting
MBHs and a small number of MBHs accreting from a thin disk as shown in
figure \ref{fig:Xray_project}.
Whether the thin disk is included or not,  MBHs
cannot account for the large number of sources at low luminosities,
which is not a problem as we expect XRBs to certainly dominate this regime.

\begin{figure*}
  \vbox{
   \includegraphics[width=15.5cm]{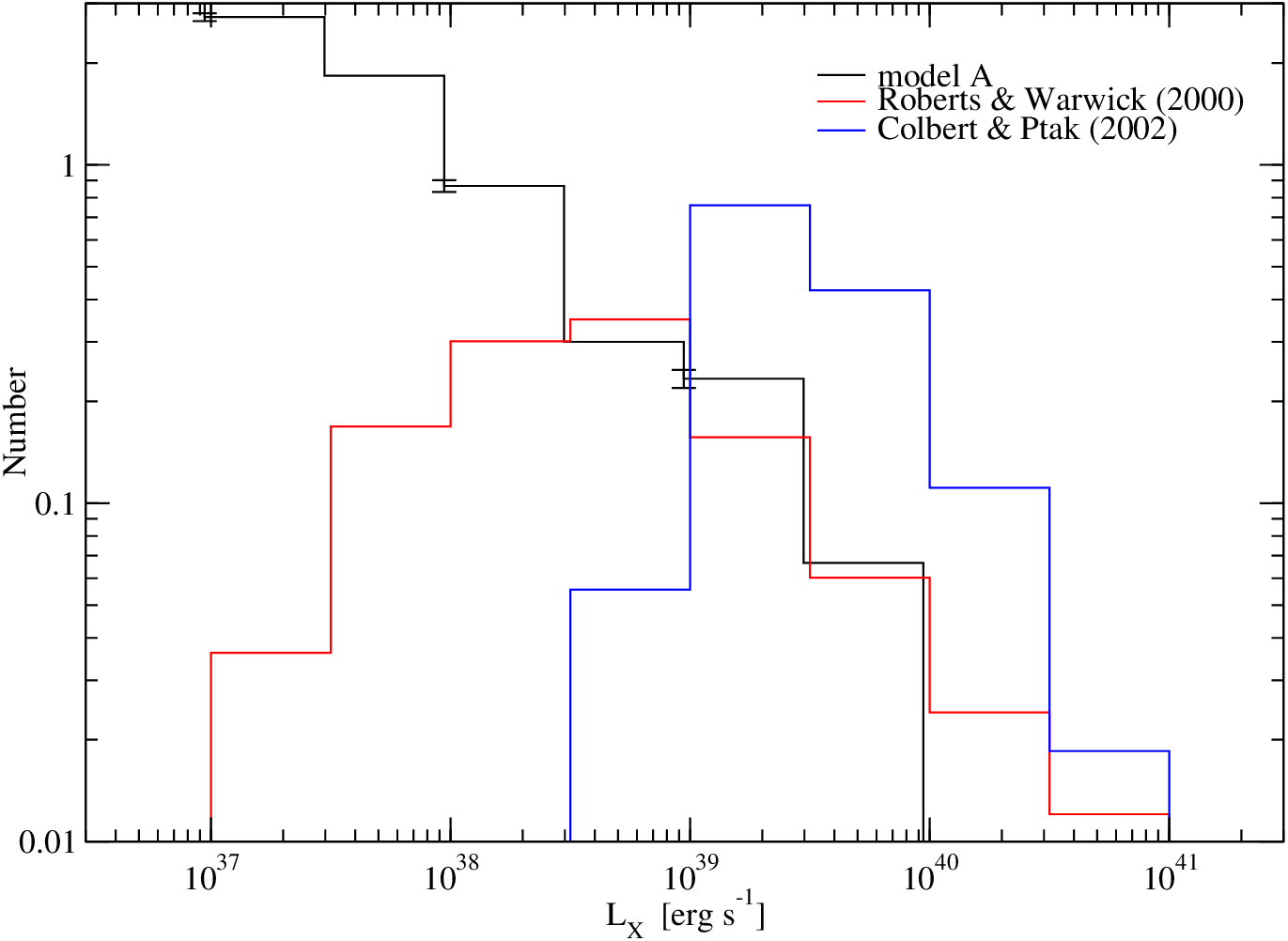}\\
    \vspace{2pt}
   \includegraphics[width=15.5cm]{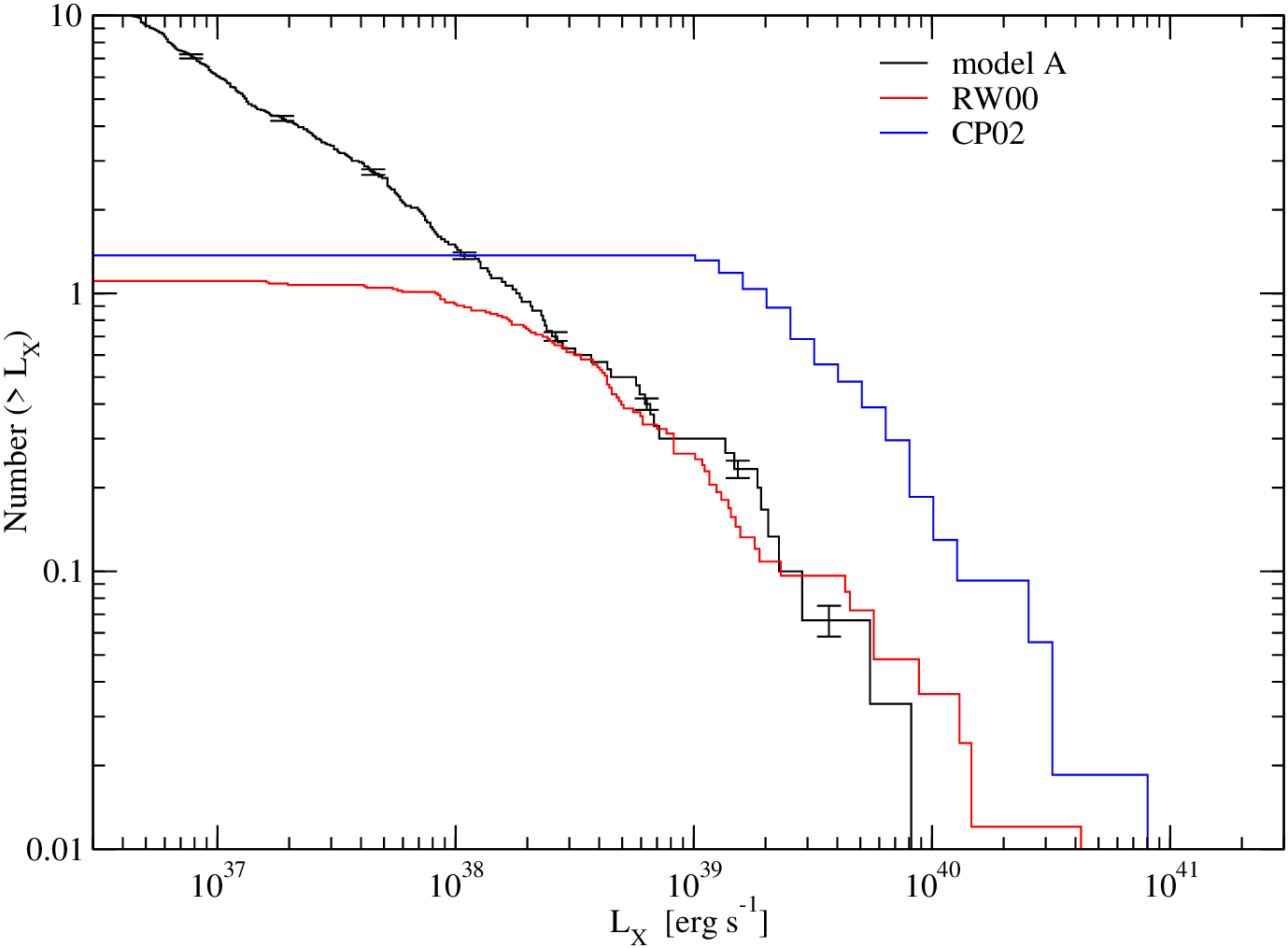}
  }
  \caption{Differential (top panel) and cumulative (bottom)
luminosity functions for
    predictions from model A ($1.6\times10^{12}$\Msun halo) and
observations. Note in particular the
    similar slope for luminosities larger than $10^{38} {\rm erg
      ~s}^{-1}$.
    Only sources at {\it radial}  distances larger than $R > 0.01 ~R_{vir}
    \approx 3$ kpc were considered. The predicted values only account
    for sources at projected distances less than $R_{proj} < 0.1 ~R_{vir}
    \approx 30$ kpc.
  }
  \label{fig:Xray_manyobservations_lum}
\end{figure*}

\begin{figure*}
  \vbox{
    \includegraphics[width=15.5cm]{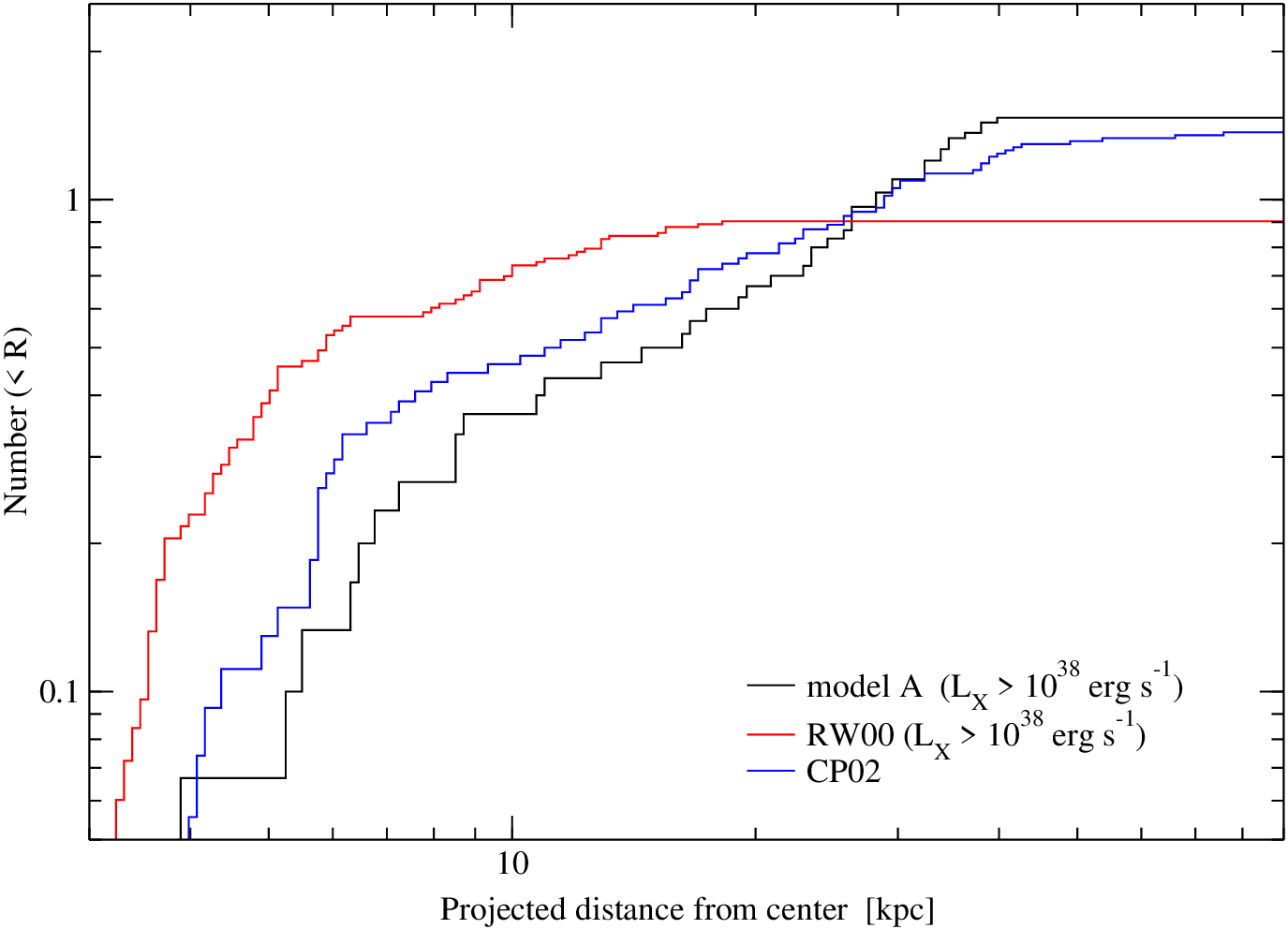}
  }
  \caption{Projected radial distribution of sources with
$L_X > 10^{38} {\rm erg~s}^{-1}$, as predicted from
    model A ($1.6\times10^{12}$\Msun halo), compared with observations.
    Only sources at {\it radial}  distances larger than $R > 0.01 ~R_{vir}
    \approx 3$ kpc were considered. The predicted values only account
    for sources at projected distances less than $R_{proj} < 0.1 ~R_{vir}
    \approx 30$ kpc.
  }
  \label{fig:Xray_manyobservations_rad}
\end{figure*}

For luminosities larger than $L_X \sim10^{38} {\rm erg ~s}^{-1}$, the
predicted average numbers of objects are near or below 1 and we have
to compare with observations of a statistical sample of galaxies.
The two data sets we use for this are that of Roberts \&Warwick (2000,
hereafter RW00) \nocite{roberts00} and Colbert \& Ptak (2002,
hereafter CP02) \nocite{colbert02}. Both sets of observations have
been obtained using the {\it ROSAT} High Resolution Imager (HRI)
instrument. In comparing with these observations, we quote the
error on the mean (and not the standard deviation of the
halo-to-halo scatter) for our predictions.
The two sets of observations vary fundamentally in that CP02 {\it only} consider galaxies that
do contain ULXs. They find a total of 87 sources in 54 galaxies of
which 15 are ellipticals. They note that in their sample the number of ULXs per galaxy
is larger for ellipticals, which account for a total of 37 ULXs. This
corresponds to 2.3 ULXs per galaxy for ellipticals compared to 1.3 per
galaxy in the case of non-ellipticals.
As the CP02 sample only counts galaxies that contain ULXs the derived
luminosity and spatial distributions will have a systematic offset
towards higher abundances.

The differential and cumulative luminosity functions are shown in
figure \ref{fig:Xray_manyobservations_lum}, for the observations and for
our expected source population for the galaxy-sized
$1.6\times10^{12}$\Msun halo in model
A. We note that the luminosity bands used in the two data sets are
different; while RW00 quote results for the {\it ROSAT} 0.1 - 2.4 keV
band, CP02 give their data for the 2.0 - 10.0 keV range.
To match these we have slightly changed our soft bands from
the one used above (0.5 - 2.0 keV). Furthermore,
we have only taken into account sources at radial distances
less than 10 \% of the halo virial radius, which corresponds to about
$R < 30$ kpc for this halo. This is also of the order of the limiting
radii below which sources in the CP02 sample are considered as
belonging to a galaxy.

Between $3\times10^{38}$ and $10^{40} {\rm erg ~ s}^{-1}$,
 our predictions in
the hard band and the soft band without the thin disk contribution
match the slope  and normalisation of RW00.
A power-law best fit yields a logarithmic slope of
$-0.83\pm0.03$,$-0.78\pm0.01$ and $-1.08\pm0.01$ for our prediction,
RW00 and CP02 respectively. For lower luminosities the
down-turn (tailing off) in the differential and cumulative
distribution of the RW00 data is probably due to incompleteness. Even
if no MBHs were present there would still be enough XRBs to match or
exceed the number of objects at lower energies.
Including the thin disk in the soft band leads to the black curve on
the right, which is ruled out even by the CP02 data which, as we
mentioned above, is biased towards higher abundances.

Aside from the uncertainty in our model parameters, there are
also several physical processes that may reduce the luminosity
of the brightest sources. The relatively large accretion rates in MBHs
accreting from thin disks could exhaust the baryonic core, if
the latter was not large enough to start with. Past merger activity of
MBH hosts might also have triggered periods of mini-quasar activity that
has lead to a faster depletion of gas in the baryonic core.
Alternately, if the accretion rates were slightly lower than assumed,
by a factor 5 or so, then most systems would fall below the critical
accretion threshold (equation \ref{eq:m_crit_ADAF}), and would accrete
in the ADAF mode. Abandoning the thin disk contribution would also
remove the brightest sources in the B and V band luminosity functions.

In figure \ref{fig:Xray_manyobservations_rad} we show the number of
sources vs projected distance from the centre. The predictions we show
are for model A and the same parameters as in figure
\ref{fig:Xray_manyobservations_lum}. We did omit the results
for the soft band without the thin disk, as this curve is
essentially the same as the one for the hard band.
The slopes of the
spatial distribution of the latter -- and by extension also that of
the soft band without thin disk -- also match that of the observations
reasonably well. However, agreement gets
increasingly worse for the RW00 sample towards small host distances
where its slope gets considerably steeper. This is probably because
RW00 have imposed a limiting minimum angular distance below which they
did not consider any sources to avoid confusion with a central
source. Just as we found no trend in the distribution of MBH masses
with radius in paper I, we do not find any indication of
`luminosity segregation' as a function of radius here.

Apart from the systematic bias in the CP02 data, the normalisation
is also affected by varying degrees of completeness in the two
samples. For instance the CP02 set features no source with $L_X <
10^{39} {\rm erg ~s}^{-1}$, and the RW00 data are restricted to much
smaller radii.
Both CP02 and RW00 maintain that contamination with background AGN is
rather small and accounts for no more than about 11 - 13 per cent of the
sample (RW00).
In the near future, larger compilations of data should
be available from {\it XMM} and {\it Chandra}.
One of the first new studies is the recent one by Swartz, Ghosh
and Tennant (2003) \nocite{swartz03b}. Data from this study are
not yet published, however.

Finally, we note that another way to probe the nature of the objects
that constitute ULXs (as well as less luminous MBH powered sources) is to
look at variability time scales of X-ray emission. A tentative linear
relation between variability time scale and BH mass has been
established although with large uncertainties \cite{markowitz03}. The
relation was found to apply to both the SMBHs powering the AGN of
Seyfert 1 galaxies as well as stellar mass galactic X-ray binary
systems. Since XRBs display variability across a wide range of time scales, this
relation, if confirmed, is probably only usefully applied to large
samples of ULXs. If both HMXRBs as well as MBHs power ULXs, variability
studies might help establish the existence of two distinct time scales
on which ULXs should be found to display variability although with large
scatter.

\subsection{Baryonic core accretion at the centre of dwarf galaxies
  and globular clusters ?}
If MBHs do accrete from residual cores of gas or stars,
then dwarf galaxies and star clusters may be likely places
to find MBHs and thus luminous X-ray point
sources. In the hierarchical structure formation scenario, dwarfs
orbiting inside the haloes of larger galaxies correspond to satellite
subhaloes, and could in principle contain accreting central MBHs.
There are roughly a dozen known dwarf galaxies within the estimated
virial radius of the Milky Way, $\sim 300$ kpc, for instance.
It would be interesting to make a systematic search for unusual
X-ray sources at the centres of these systems.

Closer to the central galaxy, a satellite is likely
to have been heavily stripped. Previously, we referred to
MBHs in these systems as `naked', although they may still be embedded
in a small baryonic core of stars or gas.
We would not expect to observe anything other than accreting MBHs
in these systems, unless the core consisted of a significantly number
of stars bound in a relatively small region.
This description appears to fit at least some globular clusters (GCs).
The observations considered above find that ULXs often appear to be
associated with GCs.
Among the formation scenarios that are considered for GCs are that
they represent remnants of high-redshift pre-galactic fragments
 \cite{beasley02}
or even that they were formed in minihaloes that gave rise to the
first massive stars \cite{lin02}. In the context of our model, both could 
correspond to the remnants of
small satellite systems that formed at high redshift. As such they are
consistent with the existence of a MBH at their centre, accreting from
a surviving baryonic core.

The lack of gas in GCs is not really an obstacle, as we already pointed
out that a baryonic core of accretable gas can in principle be very
small. What is more of a problem is the required mass for an MBH to
appear as an ULX inside a GC. The high-redshift origin of GCs implies
that at their formation MBHs only had a mass equal to the MBH
seed mass. MBHs of these masses typically only accrete
small fractions of the their Eddington accretion rate and only in few
cases this is large enough to power a ULX.
However, there is still the possibility that since then
the MBHs may have grown significantly for instance through stellar
dynamical processes inside the clusters \cite{ebisuzaki01,mouri02,miller02,zwart02}.

\subsection{Sources in and around the Milky-Way}

The Milky Way system should correspond roughly to
the $1.6\times10^{12}$ \Msun halo in our models.
For this mass we predict an average number of up to
three (model C) MBH sources with X-ray luminosities
exceeding $10^{38} {\rm erg s^{-1}}$ within the central
10 per cent of the virial radius, or $\sim$ 30 kpc.
A recent survey of X-ray sources in the
galaxy was carried out by Grimm, Gilfanov \& Sunyaev (2002)
\nocite{grimm02}.
Their work considers sources in the 2.0 - 10 keV band, which gives
probably the best representation of the sources; the optical and soft
X-ray bands are likely to be severely affected by the large
hydrogen column densities in the disk.  The
majority of the sources, and all of those above $10^{38}
{\rm erg s^{-1}}$, have been positively identified with
binary systems. The remaining unidentified sources have luminosities
significantly less than $10^{38} {\rm erg s^{-1}}$.
Even if we assume that the X-ray luminosity of MBHs is close to the
maximum in the ADAF spectrum, the absence of any non-binary ULX
in the Milky-Way is still consistent with our predictions.
On average we would expect about one to two ULXs, as can be seen in
figure \ref{fig:Xray_observations}, but the standard deviation is large
enough that many haloes would have no such sources.

\begin{figure*}
  \hbox{
    \centerline{\resizebox{\hsize}{!}{\includegraphics{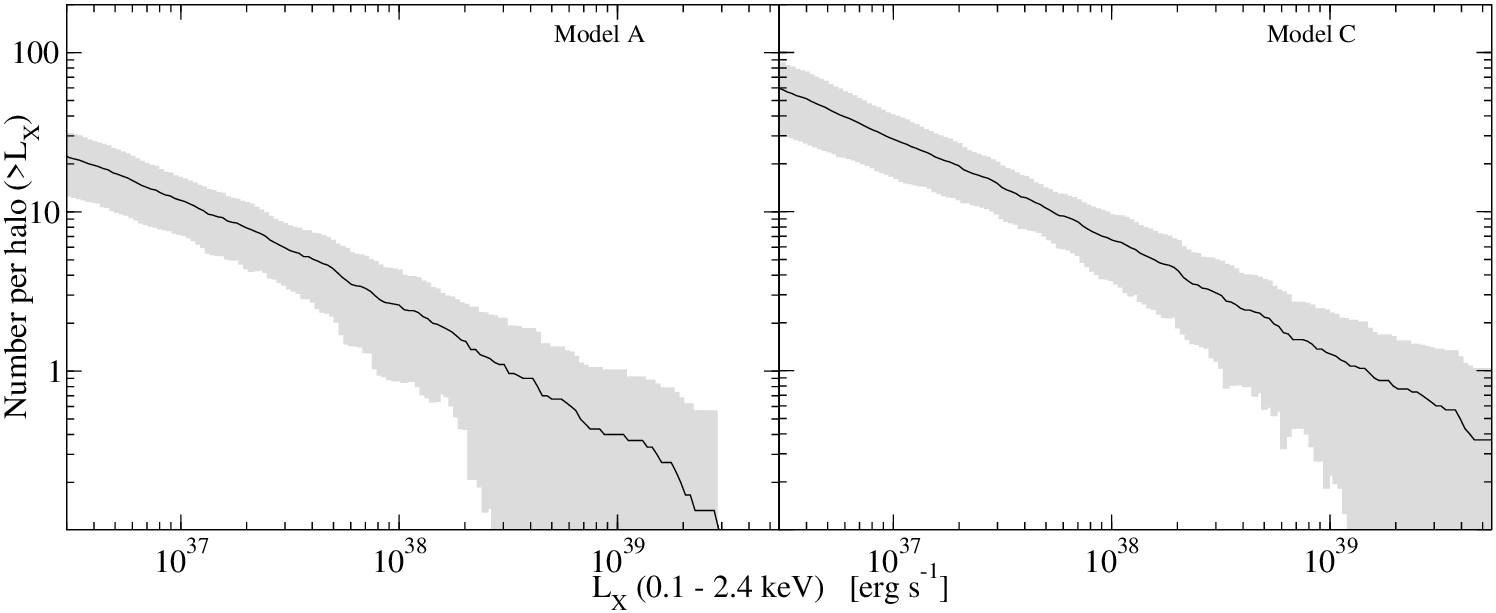}}}
  }
  \caption{Cumulative X-ray luminosity functions for a
  $1.6\times10^{12}$\Msun halo. The shaded area is the standard
  deviation around the mean number given by the lines. Model A
  is consistent with the
  absence of MBH sources more luminous than $10^{38} {\rm
  erg ~s}^{-1}$ in the Milky-Way halo, while for model C this
  represents a $\sim 1.5$-$\sigma$ deviation.
      }
  \label{fig:Milky-Waycompare}
\end{figure*}

We can also try to compare results for the entire halo of the Milky Way.
Figure  \ref{fig:Milky-Waycompare} shows the cumulative X-ray
luminosity function for the entire halo of a Milky-Way sized
galaxy. We have chosen a soft X-ray band in the energy range 0.1 - 2.4
keV to compare with observations of galaxies in the Local Group as
compiled by Zang \& Meurs (2001) \nocite{zang01}.
We are particularly interested in those Local Group galaxies that are
satellites of the Milky Way. The observations focus on the core
regions of these galaxies, which is where we would expect MBHs to
reside, if they are present at all.
For all known dwarf satellites of the Milky Way, the X-ray luminosity
from the cores does not exceed $\sim 10^{37} {\rm erg ~s}^{-1}$, and
typically it is much lower. However, for the larger satellites the
situation is different. For the Large Magellanic Cloud (LMC), for example, 
Kahabka (2002)
\nocite{kahabka02} finds a number of sources at X-ray luminosities
larger than $10^{38} {\rm erg ~s}^{-1}$ that have been classified as
binary systems. Since the classification of binaries has been done on the
basis of X-ray hardness ratios this leaves open the possibility that
some of them could be accreting MBHs.

\section{Summary and Conclusions}
\label{sec:summary}

We have examined the observational consequences of a
population of free-floating MBHs in galactic haloes.
Both our predictions and the observations of such sources
are highly uncertain. In our models, neither the mass
nor the abundance or spatial distribution of MBHs allows for a unique
identification by a single type of observation alone. As far as optical
and X-ray signatures are concerned, MBHs could be numerous enough
to differentiate them from the expected count of background sources.
On the other hand, even for optimistic choices of the model parameters,
the mass and accretion rates predicted are not large enough for the
accretion luminosity of MBHs to be of a significantly different from
that of XRBs. Forthcoming data from
{\it Chandra} and {\it XMM} should allow for better spectral modelling
of individual sources, which may resolve this ambiguity.

We have shown that our predicted sources, particularly from model A,
could account for a large fraction of the bright emitters inferred
from observations. The observed X-ray luminosity is determined by
fitting a spectral model to the detected count rate. Depending on the
model fitted the resulting luminosity can differ by a factor of order
unity, which we have deemed sufficiently accurate to compare with
predictions. This uncertainty is probably less significant than that arising
from the assumption made about the intervening absorbing column
densities, which have a strong impact on the counts of soft X-ray
photons.

However, while different spectral models may yield similar
luminosities they make significantly different assumptions about the
underlying emission process and ultimately the nature of the ULX
source. Typical fitting models include a multi-colour disk (MCD) and/or a
power law, which have the advantage of being specified by few parameters.
In most cases the colour temperature of the MCD implies a $\sim 1 - 10$
\Msun BH emitting at or above the Eddington rate.
It seems that the ADAF model applied to sub-Eddington accreting MBHs
can match not only the
X-ray luminosities but also spectral range of detected counts of ULXs.
We therefore propose to routinely include ADAF models in spectral fits
to X-ray observations.

ADAF models do involve one or two more key parameters, which may introduce
degeneracies when used to fit X-ray data alone. The advantage is that
for a single (as opposed to binary) MBH,
 ADAFs also predict the spectrum in the optical
regime and as well as the existence of a correlation of the radio and
X-ray fluxes. Consequently, observations in these bands could help break
the degeneracy
arising from the increased number of parameters. In contrast for
binaries these bands may carry the spectral contribution of the
visible companion, accounting for which adds more parameters, too.
To fit the overall spectrum ADAFs and composite
disk models for binary systems thus seem equally well placed.

Because of the absorption problem optical follow-ups are
likely to yield best results for ULXs that are observed at high
galactic latitudes to avoid the effect of the Galactic disk.
For such ULXs optical data could help distinguish between
MBHs and binary systems for sources placed within galaxies.
To test the prediction of MBHs in the dark matter haloes outside galaxies
and distinguish them from background AGN requires a systematic survey
of X-ray point sources and optical follow-ups in fields around
galaxies.\\
Our results in particular are
\begin{itemize}
\item We have applied a thin disk and ADAF to model the spectral
  signature of accreting MBHs. Applying this, we predict a few ULXs
  on average per galaxy halo. This is not consistent with
  observations, and therefore indicates either that our predicted
  accretion rates are somewhat too high, or that thin disk emission
  is shut off.
\item B and V band magnitudes reach -10, or -15 when a thin disk is
  included. Because of absorption these are probably more useful for
  follow-ups.
\item For ADAF accretion alone we match the slope of the luminosity
  function for one statistical
  sample of observations, and both, slope and
  normalisation for another one. On average we expect to find one ULX in a Milky-Way sized
  galaxy, although, the actual absence of one in the Milky-Way is consistent
  with our predictions.
\item We did not find any luminosity segregation in the MBH
  sources. In general, MBHs are distributed over a somewhat
  larger range of radii
  than observed ULXs, but this may be partly due to observational selection.
\end{itemize}

While many ULXs are clearly linked to star formation and young objects
such as high-mass XRBs, it is not clear that all ULXs need correspond
to the same physical sources. We have shown that MBHs provide an
alternative explanation for ULXs, and could plausibly account for
an appreciable fraction of observed systems, at least in terms
of the slope and normalisation of typical X-ray luminosity functions.
MBHs might be particularly interesting candidates to explain ULXs
in ellipticals, or those seen at large distances from star-forming
regions in spirals.

The accretion luminosities predicted here are depend sensitively
on the mass of MBHs and on the density of material surrounding these systems.
As explained in this paper and in paper I, we calculate these quantities
by assuming that seeds merge together efficiently, and that they
retain around them a core of roughly galactic density.
If merging does not proceed efficiently and we ignore any significant
mass increase through gas accretion, then most MBHs will have roughly
the initial seed mass and X-ray luminosities will never
reach the ULX range. Similarly, if the debris in the core around
MBHs had a much lower density then they could easily escape detection.

The luminosities of accreting MBHs in individual bands also
depend on the parameters of the particular
spectral model used. For the ADAF model, however, the predicted
X-ray luminosity is generally proportional to the bolometric
luminosity. That means we can restate our results more generally by
saying that if MBHs emit a reasonable fraction of their bolometric
luminosity in X-rays, they can account for many of the observed ULXs.
For the ADAF model the fraction of the luminosity emitted in X-rays
is predicted to be fairly high, but in general we can keep this
fraction as a free parameter encapsulating the overall uncertainty
arising from the accretion rate and the particular spectral model used.

\section*{Acknowledgements}
The authors wish to thank R.~Bandyopadhyay, G.~Bryan, J.
Magorrian and H.-W.~Rix for helpful discussions.
RRI acknowledges support from Oxford University and St.~Cross College, Oxford.
JET acknowledges support from the Leverhulme Trust and from the Particle
Physics and Astronomy Research Council (PPARC).



\begin{thebibliography}{}
\bibitem[Abel, Bryan \& Norman 2000]{abel00}
    {Abel T., Bryan G.L., Norman M.L., \ApJ{540}{39}{2000}}
\bibitem[Ball, Narayan \& Quataert 2001]
    {ball01}{Ball G.H., Narayan R., Quataert E.,
    \ApJ{552}{221}{2001}}
\bibitem[Beasley ~\ea 2002]{beasley02}
        {Beasley M.A., Baugh C.M., Forbes D.A., Sharples R.M. ~\ea,\MNRAS{333}{383}{2002}}
\bibitem[Bondi \& Hoyle 1944]{bondi44}
    {Bondi H., Hoyle F., \MNRAS{104}{273}{1944}}
\bibitem[Bondi 1952]{bondi52}
    {Bondi H., \MNRAS{112}{195}{1952}}
\bibitem[Bromm, Coppi \& Larson 2002]{bromm02}
        {Bromm V., Coppi P.S., Larson R.B., \ApJ{564}{23}{2002}}
\bibitem[Chisholm, Dodelson \& Kolb 2003]{chisholm02}
    {Chisholm J.R., Dodelson S., Kolb E.W., \ApJ{596}{437}{2003}}
\bibitem[Colbert \& Mushotzky 1999]{colbert99}
        {Colbert E.J.M., Mushotzky R.F., \ApJ{519}{89}{1999}}
\bibitem[Colbert \& Ptak 2002]{colbert02}
        {Colbert E.J.M., Ptak A.F., \ApJ{143}{S25}{2002}}
\bibitem[Ebisuzaki ~\ea 2001]{ebisuzaki01}
    {Ebisuzaki T.~\ea, \ApJ{562}{L19}{2001}}
\bibitem[Fabbiano 1989]{fabbiano89}
        {Fabbiano G., \ARAA{27}{87}{1989}}
\bibitem[Frank, King, \& Raine 1985]{frank85}
        {Frank, J., King, A.~R., \& Raine, D.~J., 1985, 
         Accretion Power in Astrophysics, Cambridge Univ.~Press, Cambridge, p.~84}
\bibitem[Fuller \& Couchman 2000]{fuller00}
        {Fuller T.M., Couchman H.M.P., \ApJ{544}{6}{2000}}
\bibitem[Grimm, Gilfanov \& Sunyaev 2002]{grimm02}
        {Grimm L.-H., Gilfanov M., Sunyaev R., \AAA{391}{923}{2002}}
\bibitem[Heger \ea 2002]{heger02}
        {Heger A., Woosley S., Baraffe I., Abel T.\ 2002, in Proc.~MPA/ESO,
        Lighthouses of the Universe: The Most Luminous Celestial Objects and 
        Their Use for Cosmology. ESO, Garching, p.~369}
\bibitem[Holt ~\ea 2003]{holt03}
        {Holt S.S., Schlegel E.M., Hwang U., Petre R., \ApJ{588}{792}{2003}}
\bibitem[Hutchings ~\ea 2002]{hutchings02}
        {Hutchings R.M., Santoro F., Thomas P.A., Couchman H.M.P.,
        \MNRAS{330}{927}{2002}}
\bibitem[Islam, Taylor \& Silk 2003, paper I herafter]{islam_I}
        {Islam R.R., Taylor J.E., Silk J., MNRAS submitted, astro-ph/0307171}
\bibitem[Ichimaru 1977]{ichimaru77}
    {Ichimaru S., \ApJ{214}{840}{1977}}
\bibitem[Ipser \& Price 1977]{ipser77}
        {Ipser J.R., Price R.H., \ApJ{216}{578}{1977}}
\bibitem[Jeltema \& Canizares 2003]{jeltema03}
        {Jeltema T.E., Canizares C.R., \ApJ{585}{756}{2003}}
\bibitem[Kaaret 2002]{kaaret02}
        {Kaaret P., \ApJ{578}{114}{2002}}
\bibitem[Kahabka 2002]{kahabka02}
        {Kahabka P., \AAA{388}{100}{2002}}
\bibitem[Kilgard ~\ea 2002]{kilgard02}
        {Kilgard, R. E., Kaaret, P., Krauss, M. I., Prestwich, A. H., Raley, M. T., Zezas, A., \ApJ{573}{138}{2002}}
\bibitem[King ~\ea 2001]{king01}
        {King, A. R., Davies, M. B., Ward, M. J., Fabbiano, G., Elvis, M., ~\ea \ApJ{552}{L109}{2001}}
\bibitem[Kormendy \& Gebhardt 2001]{kormendy01}
    {Kormendy J \& Gebhardt K., 2001,
    in Wheeler J.C., Martel H., eds, Proc.~AIP Symp.~586, XX Texas Syposium on Relativistic Astrophysics,
    AIP, New York, p.~363}
\bibitem[Larson 1981]{larson81}
        {Larson R.B, \MNRAS{194}{809}{1981}}
\bibitem[Lin \& Murray 2002]{lin02}
        {Lin D.N.C., Murray S.D., 2002, in Nomoto K., Truran J. W., eds, 
	 Proc.~IAU Symp.~Vol.~187, Cosmic Chemical Evolution. 
         Kluwer Academic, Dordrecht, p.~165}
\bibitem[Mahadevan 1997]{mahadevan97}
    {Mahadevan R., \ApJ{477}{585}{1997}}
\bibitem[Makishima ~\ea 2000]{makishima00}
        {Makishima K., Kubota A., Mizuno T., Ohnishi T., \ApJ{535}{632}{2000}}
\bibitem[Manmoto, Mineshige \& Kusunose 1997]{manmoto97}
    {Manmoto T., Mineshige S, Kusunose M., \ApJ{489}{791}{1997}}
\bibitem[Markowitz ~\ea 2003]{markowitz03}
        {Markowitz A., ~\ea, \ApJ{593}{96}{2003}}
\bibitem[Miller \& Hamilton 2002]{miller02}
    {Miller M.C., Hamilton D.P., \MNRAS{330}{232}{2002}}
\bibitem[Mouri \& Taniguchi 2002]{mouri02}
    {Mouri H., Taniguchi Y., \ApJ{566}{L17}{2002}}
\bibitem[Narayan \& Yi 1994]{narayan94}
    {Narayan R., Yi I., \ApJ{428}{L13}{1994}}
\bibitem[Narayan, Mhadevan, Quetaert 1998]{narayan98}
        {Narayan R., Mahadevan R., Quataert E., 1998, in Abramowicz M.A., 
        Bjornsson G.,Pringle J.E., eds, The Theory of
        Black Hole Accretion Disks. Cambridge Univ.~Press, Cambridge, p.~148}
\bibitem[Omukai \& Palla 2001]{omukai01}
        {Omukai K., Palla F., \ApJ{561}{L55}{2001}}
\bibitem[Portegies Zwart \& McMillan 2002]{zwart02}
    {Portegies Zwart S.F., McMillan S.L.W., \ApJ{576}{899}{2002}}
\bibitem[Rees \ea 1982]{rees82}
    {Rees M.J., Begelman M.C., Blandford R.D., Phinney E.S.,
    \Nature{295}{17}{1982}}
\bibitem[Roberts \& Warwick 2000]{roberts00}
        {Roberts T.P., Warwick R.S., \MNRAS{315}{98}{2000}}
\bibitem[Schlegel, Finkbeiner \& Davis 1998]{schlegel98}
        {Schlegel D.J., Finkbeiner D.P., Davis M. \ApJ{500}{525}{1998}}
\bibitem[Shakura \& Sunyaev 1973]{shakura73}
    {Shakura N., Sunyaev R.A., \AAA{24}{377}{1973}}
\bibitem[Smith \& Wilson 2003]{smith03}
        {Smith D.A., Wilson A.S., \ApJ{591}{138}{2003}}
\bibitem[Swartz ~\ea 2003]{swartz03}
        {Swartz D.A., Ghosh K.K., McCollough M.L., Pannuti T.G.,
        \ApJ{144}{213}{2003}}
\bibitem[Swartz, Ghosh \& Tennant 2003]{swartz03b}
        {Swartz D.A., Ghosh K.K., Tennant A.F., AAS meeting 201,
        54.13 (2003)}
\bibitem[Tegmark ~\ea 1997]{tegmark97}
        {Tegmark, M., Silk, J., Rees, M.~J., Blanchard, A., Abel, T., \& Palla, F., \ApJ{474}{1}{1997}}
\bibitem[Tennant ~\ea 2001]{tennant01}
        {Tennant, A.~F., Wu, K., Ghosh, K.~K., Kolodziejczak, J.~J., \& Swartz, D.~A., \ApJ{549}{L43}{2001}}
\bibitem[Wang 2002]{wang02}
        {Wang Q.D., \MNRAS{332}{764}{2002}}
\bibitem[Zang \& Meurs 2001]{zang01}
        {Zang Z., Meurs E.J.A., \ApJ{556}{24}{2001}}
\bibitem[Zezas \& Fabbiano 2002]{zezas02}
        {Zezas A., Fabbiano G., \ApJ{577}{726}{2002}}
\end{thebibliography}
\end{document}